\documentclass{osa-article}
\journal{oe}

\def\al{\alpha}
\def\bt{\beta}

\def\de{\delta}
\def\De{\Delta}
\def\ep{\epsilon}

\def\im{{\rm Im}}

\def\om{\omega}

\def\re{{\rm Re}}

\def\pd{\partial}

\def\si{\sigma}
\def\Si{\Sigma}
\def\sinc{{\rm sinc}}
\def\ta{\tau}

\def\sign{{\rm sign}}
\def\d{\dagger}
\def\<{\langle}
\def\>{\rangle}
\def\tint{{\textstyle\int}}

\def\d{\dagger}

\def\ba{\begin{eqnarray}}

\def\ea{\end{eqnarray}}
\def\be{\begin{equation}}
\def\ee{\end{equation}}

\begin{document}

\title{Tutorial on stochastic systems}

\author{C. J. McKinstrie, \authormark{1} T. J. Stirling \authormark{2} and A. S. Helmy \authormark{3}}

\address{\authormark{1} 3 Red Fox Run, Manalapan, New Jersey 07726, USA}

\address{\authormark{2,3} Department of Electrical and Computer Engineering, University of Toronto, Toronto, Ontario M5S 3G4, Canada}

\email{colin.mckinstrie@gmail.com}

\begin{abstract}
In this tutorial, three examples of stochastic systems are considered: A strongly-damped oscillator, a weakly-damped oscillator and an undamped oscillator (integrator) driven by noise. The evolution of these systems is characterized by the temporal correlation functions and spectral densities of their displacements, which are determined and discussed. Damped oscillators reach steady stochastic states. Their correlations are decreasing functions of the difference between the sample times and their spectra have peaks near their resonance frequencies. An undamped oscillator never reaches a steady state. Its energy increases with time and its spectrum is sharply peaked at low frequencies. The required mathematical methods and physical concepts are explained on a just-in-time basis, and some theoretical pitfalls are mentioned. The insights one gains from studies of oscillators can be applied to a wide variety of physical systems, such as atom and semiconductor lasers, which will be discussed in a subsequent tutorial.
\end{abstract}

\section{Introduction}

A wide variety of physical systems are driven by noise. Systems of interest to us include atom and semiconductor lasers \cite{sie86,mil88,agr93,col12}. To model lasers, one has to solve equations for the upper-level electon density (or number), the optical power (or photon flux) and the optical phase. These equations are nonlinear and are driven by three different noise sources. Although power and phase fluctuations have been modeled successfully, the complexity of the mathematics obscures the underlying physics, which is relatively simple. In this tutorial, three simple stochastic systems (a strongly-damped oscillator, a weakly-damped oscillator and an undamped oscillator driven by noise) are studied, with only a moderate amount of mathematics. One can apply the insights gained from this study of oscillators to the aforementioned physical systems.

This tutorial focuses on systems whose evolution is governed by stochastic differential equations of the form
\be d_t A = L(A) + R(t), \label{1.1} \ee
where $A$ is a dependent variable (an oscillator displacement or a complex wave amplitude), $L$ is a deterministic linear function of its argument and $R$ is a random function of time (noise source). In addition to linear systems, there are many nonlinear systems for which Eq. (\ref{1.1}) governs the initial growth of the amplitude, or perturbations of the steady-state amplitude. Because the amplitude is driven by noise, it too is a random function of time (or has a random component).  The properties of the amplitude are characterized by its first two moments in the time and frequency domains. The first temporal moment $M(t) = \<A(t)\>$, where $\<\ \>$ denotes an ensemble average, is called the mean, and the second moment $C(t,t') = \<A^*(t)A(t')\>$ is called the two-time correlation.
The variance $\<|A(t)|^2\> - \<A(t)\>^2$ is specified by the equal-time correlation and the mean. Similar definitions of mean, variance  and correlation apply to amplitude vectors (in which case $L$ is a matrix).

Fourier transforms are usually defined for integrable functions, which tend to zero as time tends to infinity. However, in the examples considered, noise is always present, so the driven amplitudes are not described by integrable functions. To avoid singularities in the theory, we use the finite-time transforms
\ba A(\om) &= &\int_0^T A(t) e^{i\om t} dt, \label{1.2} \\
A(t) &= &\int_{-\infty}^\infty A(\om) e^{-i\om t} d\om/2\pi. \label{1.3} \ea
In one interpretation of these equations, $A(t)$ is a Fourier series, in which the frequencies are separated by $2\pi/T$. However, if this frequency difference is much smaller than any frequency of interest, the series can be replaced by an integral. In another interpretation, $A(\om)$ is the infinite-time transform of the truncated function $A_T(t) = A(t)H(t)H(T - t)$, where $H$ is the Heaviside step function. This article is based on the latter interpretation. Although we introduced finite integration times to regularize the mathematics, they represent reality, in the sense that measurements of physical systems are taken over finite time intervals, so one can think of $T$ as the measurement time.

Some properties of Fourier transforms are reviewed in App. A. Of particular importance are the $\de$-function identity
\ba \int_{-\infty}^\infty e^{\pm i\om t} d\om/2\pi = \de(t) \label{1.4} \ea
and the Parseval equation
\be \int_0^T |A(t)|^2 dt =  \int_{-\infty}^\infty |A(\om)|^2 d\om/2\pi, \label{1.5} \ee
which follows from Eqs. (\ref{1.2}) and (\ref{1.4}).

In this article, $A(t)$ is dimensionless, so $A(\om)$ has units of time. However, in some applications (electronics and photonics), $|A(t)|^2$ has units of power, so its time integral has units of energy, and $|A(\om)|^2$ has units of energy multiplied by time (divided by frequency). For this reason, $|A(\om)|^2$ is called the spectral energy density and $|A(\om)|^2/T$ is called the spectral power density. We will use the descriptive terms power and energy whenever it is helpful to do so. The Parseval equation states that the time- and frequency-domain formulas for the energy of a system are equivalent. This equivalence provides a way to check the consistency of time- and frequency-domain analyses.

The two-frequency correlation $C(\om,\om') = \<A^*(\om)A(\om')\>$, where $A^*(\om)$ is an abbreviation for $[A(\om)]^*$, is the double Fourier transform of the two-time correlation $C(t,t')$. In turn, the temporal correlation is the double inverse transform of the frequency correlation. Some consequences of this relation are discussed in App. B. In the main text, it is sufficient to consider the spectral energy density $S(\om) = C(\om,\om)$, which is specifed by the equation
\be S(\om)  = \int_0^T \int_0^T C(t,t') e^{i\om(t' - t)} dtdt'. \label{1.6} \ee
Thus, if the temporal correlation is known, the spectrum can be calculated (at least in principle).

This tutorial is organized as follows: In Sec. 2, the properties of white and colored noise are defined and discussed. In Secs. 3 -- 5, the behaviors of strongly-damped, weakly-damped and undamped oscillators driven by noise are studied. Damped oscillators attain stochastic steady states, whereas undamped oscillators do not. Formulas are derived for the temporal correlations of the displacements and their associated spectra. These formulas are nonsingular, but only because the integration (measurement) time is nonsingular. In Sec. 6, the specific results obtained in the preceding sections are discussed in the broader context of stochastic processes \cite{gar85,goo85,pap91,lem02}. Finally, in Sec. 7, the main results of this tutorial are summarized. For convenience, the supplemental material of this tutorial contains appendices on Fourier transforms and the Wiener--Khinchin theorem \cite{wie30,khi34}.

\section{Noise}

The most common model of noise is white noise, which is defined by the properties
\ba \<R(t)\> &= &0, \label{2.1} \\
\<R(t)R(t')\> &= &0, \label{2.2} \\
\<R^*(t)R(t')\> &= &\si\de(t - t'), \label{2.3} \ea
where $\si$ is the source strength. $R(t)dt$ is the impulse applied to a realization of the system during the time interval $dt$. The first equation states that the average impulse is zero (positive and negative, real and imaginary impulses are equally likely). The second and third equations state that different impulses are independent, and the real and imaginary parts of the same impulse have the same strength ($\si/2$), but  are otherwise independent.

Although this noise model is common, the assumption of zero correlation time is problematic. First, the noise power $\<|R(t)|^2\> = \si\de(0)$ is undefined (infinite), as is the associated noise energy. Second, the spectral energy density
\ba \<|R(\om)|^2\> &= &\int_0^T \int_0^T \<R^*(t)R(t')\> e^{-i\om(t - t')} dt dt' \nonumber \\
&= &\int_0^T \si dt \ = \ \si T. \label{2.4} \ea
The spectral density is finite, because the integration time is finite, but the spectral energy, which is the frequency integral of the density, is still infinite.

One can resolve these problems by assuming that the correlation time is very short, but nonzero. Colored noise also has properties (\ref{2.1}) and (\ref{2.2}). The correlation function
\be \<R^*(t)R(t')\> = (\si\nu/2) e^{-\nu|t - t'|} \label{2.5} \ee
is exponentially localized and has a correlation (coherence) time of $1/\nu$. (Other localized functions are also suitable.) The normalized correlation, which is correlation (\ref{2.5}) divided by $(\si\nu/2)$, is plotted as a function of the time difference $\ta = t' - t$ in Fig. 1(a). The noise power $\<|R(t)|^2\> = \si\nu/2$ is finite, because $\nu$ is finite, and the noise energy
\be \int_0^T |R(t)|^2 dt = \int_0^T (\si\nu/2) dt = \si\nu T/2 \label{2.6} \ee
is finite, because $\nu$ and $T$ are both finite.
\begin{figure}[h!]
\vspace{-0.15in}
\centerline{\includegraphics[width=2.6in]{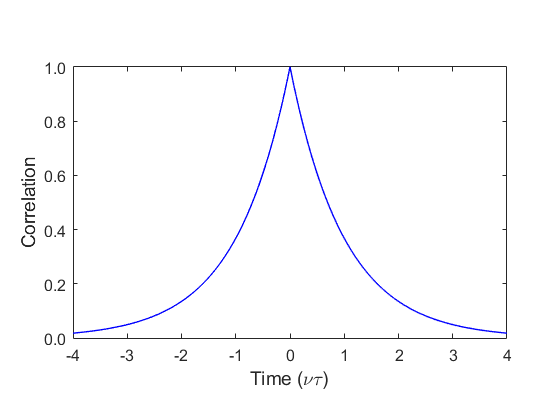} \hspace{0.0in} \includegraphics[width=2.6in]{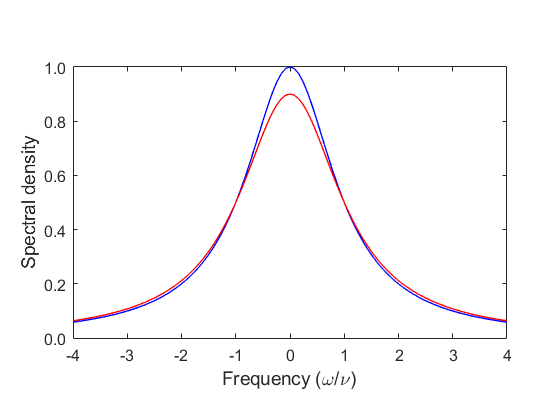}}
\vspace{-0.1in}
\caption{Temporal correlation function (left) and spectral energy density (right) of colored noise. The blue density curve represents the first term in Eq. (\ref{2.10}), whereas the red curve denotes both terms. For $\nu T = 10$, the finite-time correction is noticible.}
\end{figure}

The spectral energy density
\be \<|R(\om)|^2\> = \int_0^T \int_0^T (\si\nu/2) e^{-\nu|t - t'|} e^{i\om(t' - t)} dt dt'. \label{2.7} \ee
On the right side of Eq. (\ref{2.7}), the $t'$-integral is
\ba &&\int_0^t e^{-(\nu + i\om)(t - t')} dt' + \int_t^T e^{-(\nu - i\om)(t' - t)} dt' \nonumber \\
&= &\frac{1 - e^{-(\nu + i\om)t}}{\nu + i\om} + \frac{1 - e^{-(\nu - i\om)(T-t)}}{\nu - i\om}. \label{2.8} \ea
The constant terms in Eq. (\ref{2.8}) sum to $2\nu/(\nu^2 + \om^2)$, the $t$-integral of which is $2\nu T/(\nu^2 + \om^2)$. The $t$-integral of the variable terms is
\be -\frac{1 - e^{-(\nu + i\om)T}}{(\nu + i\om)^2} - \frac{1 - e^{-(\nu - i\om)T}}{(\nu - i\om)^2}. \label{2.9} \ee
The terms proportional to $e^{-\nu T}$ are exponentially small and the other terms sum to $-2(\nu^2 - \om^2)/(\nu^2 + \om^2)^2$. This contribution is smaller than the previous one by a factor of order $\nu T$ and its frequency integral is zero (as are the integrals of the exponential terms). By combining the preceding results, one obtains the spectrum
\be \<|R(\om)|^2\> = \si\nu \Biggl[\frac{\nu T}{\nu^2 + \om^2} - \frac{\nu^2 - \om^2} {(\nu^2 + \om^2)^2}\Biggr]
\sim \frac{\si\nu^2T}{\nu^2 + \om^2}. \label{2.10} \ee
The normalized spectral density, which is density (\ref{2.10}) divided by $\si T$, is plotted as a function of frequency in Fig. 1(b). Notice that spectrum (\ref{2.4}) is constant, whereas spectrum (\ref{2.10}) varies with frequency, which is why formulas (\ref{2.3}) and (\ref{2.5}) are said to define white and colored noise, respectively.
The frequency integral of the Lorentzian function $1/(\nu^2 + \om^2)$ is $\pi/\nu$, so the integrated spectrum
\be \int_{-\infty}^\infty \<|R(\om)|^2\> d\om/2\pi = \si\nu T/2. \label{2.11} \ee
Equation (\ref{2.11}) is identical to Eq. (\ref{2.6}). For colored noise with an exponential correlation function, the spectrum is (asymptotically) Lorentzian and the Parseval relation is satisfied.

Notice that for white noise, the spectral power density $\<|R(\om)|^2\>/T = \si$, which is the two-sided-in-time Fourier transform of correlation function (\ref{2.3}) with respect to the time difference $t - t'$. For colored noise, $\<|R(\om)|^2\>/T \sim \si\nu^2/(\nu^2 + \om^2)$, which is the transform of correlation function (\ref{2.5}) with respect to the time difference. These results are not coincidental. They are manifestations of the Wiener--Khinchin theorem, which states that if the correlation function depends on the time difference $t - t'$ (rather than $t$ and $t'$ separately), then the spectral power density is (asymptotically) equal to the two-sided Fourier transform of the correlation function with respect to the time difference. This important result is proved in App. C.

In summary of this section, the energy provided by a noise source is only finite if the measurement (reception) time and the noise bandwidth are both finite. If the correlation time of colored noise is much shorter than any other time scale relevant to the receiving system, its effects on that system should be similar to those of white noise. For this reason, one usually analyses the effects of white noise first, because the analysis is simpler, and checks the results for singularities. If none are present, then the white-noise model is sufficient.

\section{Strongly-damped oscillator}

Consider the response of a driven oscillator (or wave), which is governed by the displacement (or amplitude) equation
\be d_t A = -(\nu_r + i\nu_i)A + R_a(t), \label{3.1} \ee
where $\nu_r$ and $\nu_i$ are damping and detuning (frequency-shift) parameters, respectively. The Langevin source term $R_a$ \cite{lem97} is a random function of time with the properties (\ref{2.1}) -- (\ref{2.3}).
For example, Eq. (\ref{3.1}) governs the evolution of a cavity-mode amplitude in a below-threshold laser.
Let $A(t) = B(t)\exp(-i\nu_it)$. Then the transformed amplitude satisfies the stochastic differential equation
\be d_tB = -\nu_rB + R_b(t), \label{3.2} \ee
where $R_b(t) = R_a(t)\exp(i\nu_it)$. It is easy to verify that $R_b$ has the same properties as $R_a$. In statistical physics, Eq. (\ref{3.2}) defines an Ornstein--Uhlenbeck process \cite{gar85,lem02}.

Equation (\ref{3.2}) is linear in the dependent variable, so one can use the superposition principle to write its solution in the Green-function form
\be B(t) = \int_0^t G(t - s)R(s) ds, \label{3.3} \ee
where $G(t)$ is the solution of the equation
\be d_t G = -\nu_r G + \de(t), \label{3.4} \ee
together with the initial condition $G(0) = 0$. The solution of Eq. (\ref{3.4}) is $H(t)e^{-\nu_rt}$, from which it follows that the solution of Eq. (\ref{3.3}) is
\be B(t) = \int_0^t e^{-\nu(t - s)} R(s) ds, \label{3.5} \ee
where $R$ and $\nu$ are abbreviations of $R_b$ and $\nu_r$, respectively. Although the impulse $R(s)ds$ is a random function of $s$, its subsequent effect on the amplitude is determined by a homogenous linear equation, which one can solve using regular calculus. The stochastic differential equations that will be considered in Secs. 4 and 5 are also linear, so regular calculus will be used throughout this tutorial. For nonlinear equations driven by noise (such as those that govern laser power and phase fluctuations), one has to choose between Ito and Stratonovich calculus \cite{gar85}. It is fortunate that some interesting stochastic systems can be discussed without the extra complexity of stochastic calculus.

It follows from solution (\ref{3.3}) and property (\ref{2.1}) that $\<B(t)\> = 0$. The amplitude is a random function of time, with a mean of zero. The amplitude variance $V(t) = \<|B(t)|^2\> - \<|B(t)|\>^2$. For systems such as this one, which have means of zero, the variance equals the power $P(t) = \<|B(t)|^2\>$. It follows from solution (\ref{3.3}) that
\ba \<|B(t)|^2\> &= &\int_0^t \int_0^t \<R^*(s)R(s')\> e^{-\nu(2t - s - s')} dsds' \nonumber \\
&= &\si \int_0^t e^{-2\nu(t - s)} ds \nonumber \\
&= &\si (1 - e^{-2\nu t})/2\nu. \label{3.6} \ea
Initially, the power is zero, because the oscillator starts from rest. Subsequently, the power increases monotonically with time. For times longer than the damping time, it tends to its steady-state value $\si/2\nu$, which is proportional to the source strength and inversely proportional to the damping rate. The steady-state power is finite, because the damping rate is nonzero. The energy
\be \int_0^T \<|B(t)|^2\> dt = \si (T/2\nu - 1/4\nu^2), \label{3.7} \ee
where a term proportional to $e^{-2\nu T}$ was neglected. The energy is finite because the damping rate is nonzero and the measurement time is finite. The second term on the right side of Eq. (\ref{3.7}), which is associated with the transient power deficit, is smaller than the first term by a factor of $2\nu T$. Although this energy deficit is small, it is noticible (not exponentially small).

The two-time correlation function
\be \<B^*(t)B(t')\> = \int_0^t \int_0^{t'} \<R^*(s)R(s')\> e^{-\nu(t + t' - s - s')} ds'ds. \label{3.8} \ee
It follows from the Cauchy--Schwartz inequality that the magnitude of the correlation is less than or equal to the power. Because the source function is $\de$-correlated in time, the noise applied between $t$ and $t'$ does not contribute to the correlation, so one can replace the upper limits of integration by $\min(t,t')$. Suppose that $t' > t$. Then
\ba \<B^*(t)B(t')\> &= &\si \int_0^t e^{-\nu(t + t' - 2s)} ds \nonumber \\
&= &\si [e^{-\nu(t' - t)} - e^{-\nu(t' + t)}]/2\nu. \label{3.9} \ea
Because correlation (\ref{3.9}) is real, it must be a symmetric function of $t$ and $t'$. One can make it so by replacing $t' - t$ with $|t' - t|$. Correlation (\ref{3.9}) is illustrated in Fig. 2. The ridge on the main diagonal represents strong correlations.
\begin{figure}[h!]
\vspace{-0.1in}
\centerline{\includegraphics[width=2.8in]{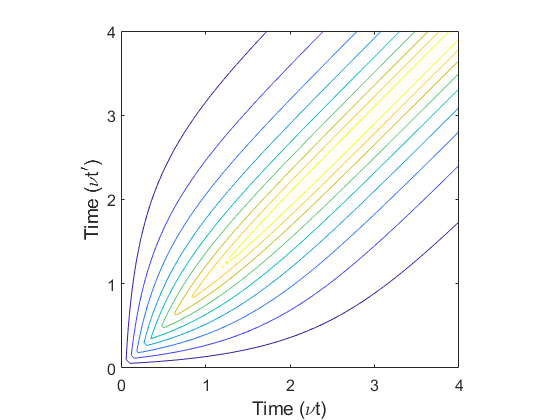}}
\vspace{-0.0in}
\caption{Two-time correlation function of a strongly-damped oscillator. For times that are longer than the damping time, the correlation depends on only the time difference $t' - t$.}
\end{figure}

For times ($t$ and $t'$) that are longer than the damping time, the second term on the right side of Eq. (\ref{3.9}) is negligible and the correlation is a function of only the time difference $\ta = t' - t$. The correlation does not depend on the time origin, so $C(t,t+\ta) = C(0,\ta)$. In abbreviated notation,
\be C(\ta) \sim \si e^{-\nu|\ta|}/2\nu. \label{3.10} \ee
The normalized correlation, which is correlation (\ref{3.10}) divided by $\si/2\nu$,  is plotted as a function of the time difference in Fig. 3(a). It decreases exponentially with the difference (positive or negative). This behavior reflects the fact that each noise impulse only affects the cumulative displacement for a time of the order of the damping time [Eq. (\ref{3.4})].
\begin{figure}[h]
\vspace{-0.1in}
\centerline{\includegraphics[width=2.6in]{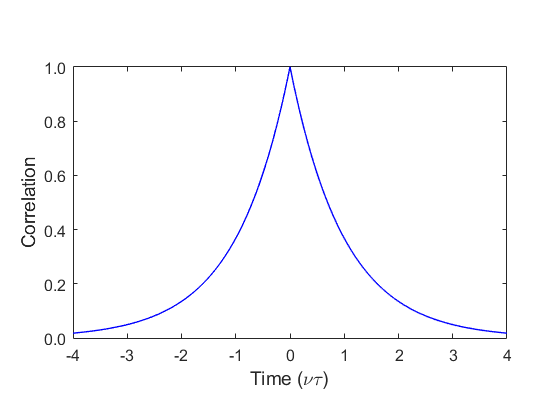} \hspace{0.0in} \includegraphics[width=2.6in]{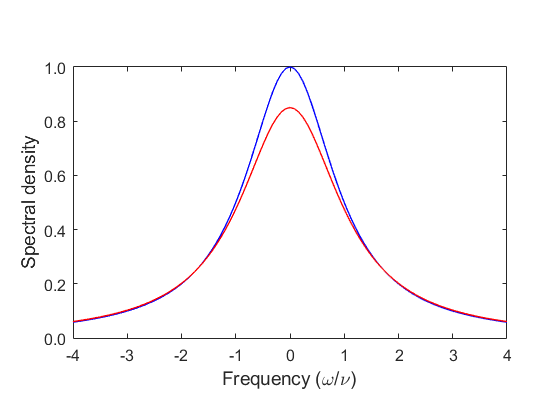}}
\vspace{-0.1in}
\caption{Temporal correlation function (left) and spectral energy density (right) of a strongly-damped oscillator. The blue density curve represents the first term in Eq. (\ref{3.11}), whereas the red curve denotes all three terms. For $\nu T = 10$, the finite-time and transient corrections are noticible.}
\end{figure}

The correlation function (\ref{3.9}) has two contributions: The first (longer-lasting) term is proportional to the correlation function of colored noise, which was specified in Eq. (\ref{2.5}). Its contribution to the spectrum was stated in Eq. (\ref{2.10}). The contribution to the spectrum of the second (transient) term is $-\si/2\nu(\nu^2 + \om^2)$, plus exponentially small terms. Hence, the spectrum of a strongly-damped oscillator is
\be \<|B(\om)|^2\> = \frac{\si T}{\nu^2 + \om^2} - \frac{\si(\nu^2 - \om^2)}{\nu(\nu^2 + \om^2)^2} - \frac{\si}{2\nu(\nu^2 + \om^2)}. \label{3.11} \ee
The normalized spectral density, which is density (\ref{3.11}) divided by $\si T/\nu^2$, is plotted as a function of frequency in Fig. 3(b).
This spectrum is also asymptotically Lorentzian. Although the transient contribution (deficit) is small, it is noticible. [Compare Figs. 1(b) and 3(b).] The integral of the second term on the right side of Eq. (\ref{3.11}) is 0 and the integral of $1/(\nu^2 + \om^2)$ is $\pi/\nu$. Hence, the spectral energy
\be \int_{-\infty}^\infty \<|B(\om)|^2\> d\om/2\pi = \si(T/2\nu - 1/4\nu^2). \label{3.12} \ee
Equation (\ref{3.12}) is consistent with Eq. (\ref{3.7}). Although the noise spectral energy is infinite (Sec. 2), the displacement spectral energy is finite. Damping acts like a frequency filter, accepting only noise components with frequencies between (roughly) $-2\nu$ and $2\nu$. In doing so, it converts white noise to colored noise.

One can also work in the frequency domain, in which the source function has the properties
\ba \<R(\om)\> &= &\int_0^T \<R(t)\> e^{i\om t} dt \ = \ 0, \label{3.13} \\
\<|R(\om)|^2\> &= &\int_0^T \int_0^T \<R^*(t)R(t')\> e^{i\om(t' - t)} dtdt' \ = \ \si T. \label{3.14} \ea
Other properties of the source function are described in App. B. It is a basic property of Fourier transforms that
\be \int_0^T [d_t B(t)] e^{i\om t} dt = B(T)e^{i\om T} - B(0) - i\om \int_0^T B(t) e^{i\om t} dt. \label{3.15} \ee
In this article, the oscillators start from rest, so $B(0) = 0$. If they were not driven, then their responses would be transient and $B(T)$ would tend to 0 as $t \rightarrow \infty$. However, they are driven by noise, which is always present, so $B(T) \neq 0$ and one should apply the rule that the Fourier transform of $d_tB$ is the transform of $B$ multiplied by $-i\om$ with caution.

By transforming Eq. (\ref{3.2}), applying the initial condition $B(0) = 0$ and omitting $B(T)$, one finds that
\be B(\om) = R(\om)/(\nu - i\om). \label{3.16} \ee
The convolution theorem (App. A) shows that the frequency-domain solution (\ref{3.16}) is consistent with the time-domain solution (\ref{3.5}). [In App. D, the inverse transform of $1/(\nu - i\om)$ is shown to be $H(t)e^{-\nu t}$.] By combining Eqs. (\ref{3.14}) and (\ref{3.16}), one obtains the spectrum
\be \<|B(\om)|^2\> = \si T/(\nu^2 + \om^2). \label{3.17} \ee
The spectrum in Eq. (\ref{3.17}) is consistent with the first term in Eq. (\ref{3.11}). It misses the second and third terms, which are of relative order $1/\nu T$. (The second term is the finite-time correction to the Lorentzian and the third term is the transient deficit.) Thus, for a strongly-damped oscillator, which attains a steady stochastic state, the standard frequency-domain analysis, which omits $B(T)$, is only asymptotically correct.

Results (\ref{3.10}) and (\ref{3.11}) apply to the transformed amplitude $B$. For the amplitude $A$, the corresponding results are
\ba \<A^*(0)A(\ta)\> &\sim &(\si/2\nu_r)e^{-\nu_r|\ta| - i\nu_i\ta}, \label{3.18} \\
\<|A(\om)|^2\>/T &\sim &\si/[\nu_r^2 + (\om - \nu_i)^2], \label{3.19} \ea
respectively. Although the imaginary part of the rate coefficient shifts the phase of the correlation and the peak of the spectrum, the duration of the correlation and width of the spectrum depend on only the real part of the coefficient. Notice that formulas (\ref{3.18}) and (\ref{3.19}) satisfy the Wiener--Khinchin theorem.

\section{Weakly-damped oscillator}

In this section, the response of a weakly-damped oscillator to a noise source is studied. The material is important for two reasons. First, oscillatory behavior is ubiquitous in science and engineering. Second, the response of a weakly-damped oscillator is intermediate between the responses of a strongly-damped oscillator (Sec. 3) and an undamped oscillator (Sec. 5). However, the physical results of this section are similar to those of Sec. 3 (after transients have died out, the oscillator displacement reaches a stochastic steady state in which the temporal correlation depends on only the time difference and the spectrum is peaked near the natural frequency), but require much more mathematical effort to obtain. Consequently, we recommend that readers proceed directly to Sec. 5 and return to Sec. 4 later.

Let $X$ and $V$ be the oscillator displacement and velocity, respectively. Then the oscillator equations are
\ba d_t X &= &V, \label{4.1} \\
d_t V &= &-2\nu_0 V - \om_0^2 X + \ep R(t), \label{4.2} \ea
where $\nu_0$ and $\om_0$ are damping and frequency parameters, respectively. In Eq. (\ref{4.2}), $\ep$ is a book-keeping parameter with units of frequency, which allows the source function $R$ to have the same properties in this section as it did in the previous ones [Eqs. (\ref{2.1}) -- (\ref{2.3})]. The oscillator equations are similar to the ones that govern electron-number ($N$) and photon-number ($P$) fluctuations in an above-threshold laser \cite{agr93,col12}. $X$ corresponds to $P$ and $V$ corresponds to $N$.

Equations (\ref{4.1}) and (\ref{4.2}) are linear, so their solutions can be written in the forms
\ba X(t) &= &\ep \int_0^t G_x(t - s) R(s) ds, \label{4.3} \\
V(t) &= &\ep \int_0^t G_v(t - s) R(s) ds, \label{4.4} \ea
where the time-domain Green functions
\ba G_x(t) &= &H(t)\sin(\om_rt)e^{-\nu_0t}/\om_r, \label{4.5} \\
G_v(t) &= &H(t)[\cos(\om_rt) - \nu_0\sin(\om_rt)/\om_r]e^{-\nu_0t}, \label{4.6} \ea
and the resonance frequency $\om_r = (\om_0^2 - \nu_0^2)^{1/2}$. Both response functions decrease with time, but in an oscillatory manner (rather than a monotonic one). Notice that $G_v = d_t G_x$ for positive times (as it should do). 

It follows from Eqs. (\ref{2.3}), (\ref{4.3}) and (\ref{4.5}) that the squared displacement (power) is
\ba \<X^2(t)\> &= &\ep^2\si \int_0^t G_x^2(t - s) ds \nonumber \\
&= &(\ep^2\si/2\om_r^2) \int_0^t [1 - \cos(2\om_r s)]e^{-2\nu_0 s} ds. \label{4.11} \ea
By integrating by parts twice, one can derive the identities
\ba \int_a^b \cos(\al s)e^{-\bt s} ds &=&\frac{[\al\sin(\al s) - \bt\cos(\al s)]e^{-\bt s} |_a^b}{\al^2 + \bt^2}, \label{4.12} \\
\int_a^b \sin(\al s)e^{-\bt s} ds &= &-\frac{[\al\cos(\al s) + \bt\sin(\al s)]e^{-\bt s} |_a^b}{\al^2 + \bt^2}. \label{4.13} \ea
Notice that if $b$ is large (or infinite), then only the $a$-contributions are significant. Notice also that if $\al = 2\om_r$ and $\bt = 2\nu_0$, then $\al^2 + \bt^2 = 4\om_0^2$.
By combining Eqs. (\ref{4.11}) and (\ref{4.12}), one finds that
\ba \<X^2(t)\> &= &\frac{\ep^2\si}{2\om_r^2} \Bigg\{ \frac{1 - e^{-2\nu_0 t}}{2\nu_0} - \frac{\om_r \sin(2\om_r t)e^{-2\nu_0 t}}{2\om_0^2} \nonumber \\
&&-\ \frac{\nu_0[1 - \cos(2\om_r t)e^{-2\nu_0 t}]}{2\om_0^2} \Bigg\}. \label{4.14} \ea
Initially, the power is zero. After a time of the order of the damping time $1/\nu_0$, the transient contributions die out and the power attains its steady-state value
\be \<X^2\> = \frac{\ep^2\si}{2\om_r^2}\Biggl(\frac{1}{2\nu_0} - \frac{\nu_0}{2\om_0^2}\Biggr)
= \frac{\ep^2\si}{4\nu_0\om_0^2}. \label{4.15} \ee

The energy is the time integral of the power. The components integrals are
\ba (1/2\nu_0) \int_0^T (1 - e^{-2\nu_0t}) dt &= &T/2\nu_0 - 1/4\nu_0^2, \label{4.16} \\
(\om_r/2\om_0^2) \int_0^T \sin(2\om_r t)e^{-2\nu_0 t} dt &= &(\om_r/2\om_0^2)^2, \label{4.17} \\
(\nu_0/2\om_0^2) \int_0^T [1 - \cos(2\om_r t)e^{-2\nu_0 t}] dt &= &(\nu_0/2\om_0^2)(T - \nu_0/2\om_0^2), \label{4.18} \ea
where terms proportional to $e^{-2\nu_0T}$ were omitted. In Eq. (\ref{4.16}), the transient contribution is smaller than the steady-state contribution by the factor $2\nu_0T$, and in Eq. (\ref{4.18}) it is smaller by the factor $2\om_0^2T/\nu_0 = 2\nu_0T(\om_0/\nu_0)^2$. It follows from these results that the energy
\be \int_0^T \<X^2(t)\> dt \sim \frac{\ep^2\si T}{2\om_r^2} \Biggl( \frac{1}{2\nu_0} - \frac{\nu_0}{2\om_0^2}\Biggr) = \frac{\ep^2\si T}{4\nu_0\om_0^2}, \label{4.19} \ee
which is just the steady-state power multiplied by the measurement time.

It also follows from Eqs. (\ref{2.3}), (\ref{4.3}) and (\ref{4.5}) that the temporal correlation
\ba \<X(t)X(t')\> &= &\ep^2\si \int_0^t G_x(t - s)G_x(t' - s) ds \nonumber \\
&= &(\ep^2\si/2\om_r^2) \int_0^t \{\cos[\om_r(t'-t)] \nonumber \\
&&-\ \cos[\om_r(t'+t-2s)]\}e^{-\nu_0(t'+t-2s)} ds. \label{4.21} \ea
By making the substitution $u = t + t' - 2s$, one can rewrite the first integral in Eq. (\ref{4.21}) as
\ba I_1(t,t') &= &\cos[\om_r(t' - t)] \int_{t'-t}^{t'+t} e^{-\nu_0u} du/2 \nonumber \\
&= &\cos[\om_r(t' - t)][e^{-\nu_0(t' - t)} - e^{-\nu_0(t'+t)}]/2\nu_0. \label{4.22} \ea
By repeating this substitution and using identity (\ref{4.12}), one can rewrite the second integral as
\ba I_2(t,t') &= &\int_{t'-t}^{t+t'} \cos(\om_ru)e^{-\nu_0u} du/2 \nonumber \\
&= &\{\om_r\sin[\om_r(t'+t)] - \nu_0\cos[\om_r(t'+t)]\} \nonumber \\
&&\times\ e^{-\nu_0(t'+t)}/2\om_0^2 \nonumber \\
&&-\ \{\om_r\sin[\om_r(t'-t)] - \nu_0\cos[\om_r(t'-t)]\} \nonumber \\
&&\times\ e^{-\nu_0(t'-t)}/2\om_0^2. \label{4.23} \ea
It follows from Eqs. (\ref{4.21}) -- (\ref{4.23}) that the correlation
\be C(t,t') = (\ep^2\si/2\om_r^2) [I_1(t,t') - I_2(t,t')]. \label{4.24} \ee
Formula (\ref{4.24}) was derived for positive time differences. Because the correlation is real, it must be a symmetic function of $t$ and $t'$. One can make it so by replacing $t' - t$ with $|t' - t|$ in Eqs. (\ref{4.22}) and (\ref{4.23}). Correlation (\ref{4.24}) is illustrated in Fig. 4. Notice that the diagonal ridge is flanked by two ravines.
\begin{figure}[h!]
\vspace{-0.1in}
\centerline{\includegraphics[width=2.8in]{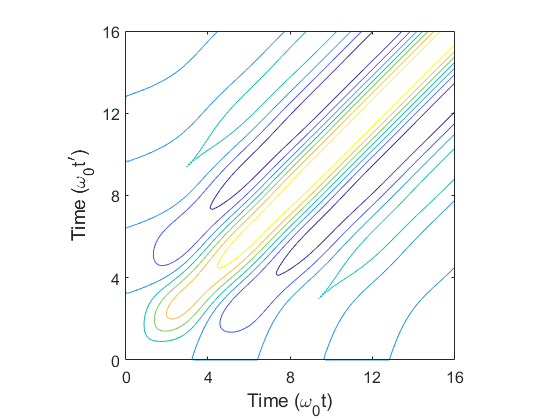}}
\vspace{-0.0in}
\caption{Two-time correlation function of a weakly-damped oscillator for $\nu_0/\om_0 = 0.2$. For times that are longer than the damping time, the correlation depends on only the time difference $t' - t$.}
\end{figure}

For times ($t$ and $t'$) that are longer than the damping time $1/\nu_0$, the second contributions to integrals (\ref{4.22}) and (\ref{4.23}) can be neglected. By omitting these terms and combining the cosine terms that remain, one obtains the time-asymptotic correlation
\be C(\ta) \sim \frac{\ep^2\si}{4\nu_0\om_0^2} \biggl[\cos(\om_r\ta) + \frac{\nu_0}{\om_r}\sin(\om_r|\ta|)\biggr]e^{-\nu_0|\ta|}, \label{4.25} \ee
where $\ta = t' - t$. The normalized correlation, which is correlation (\ref{4.25}) divided by $(\ep^2\si/4\om_0^2)$, is plotted as a function of the time difference in Fig. 5(a).
\begin{figure}[h]
\vspace*{-0.1in}
\centerline{\includegraphics[width=2.6in]{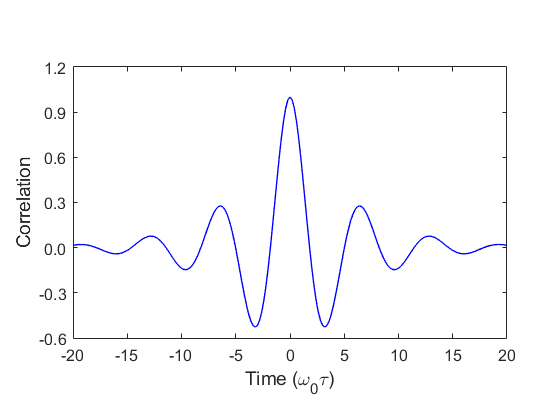} \hspace{0.0in} \includegraphics[width=2.6in]{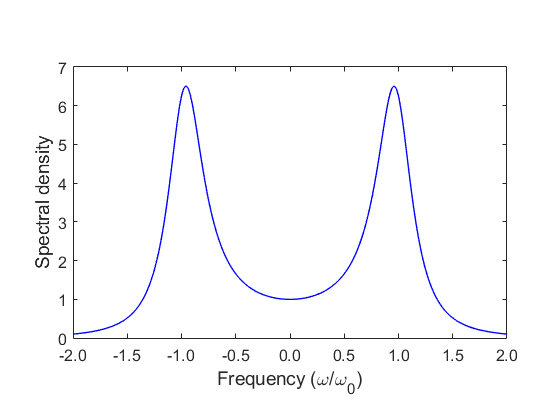}}
\vspace*{-0.1in}
\caption{Temporal correlation function (left) and spectral energy density (right) of a weakly-damped oscillator for $\nu_0/\om_0 = 0.2$.}
\end{figure}

The spectral energy density is defined by Eqs. (\ref{1.6}) and (\ref{4.25}), the second of which can be rewritten in the form
\be C(\ta) \propto [(\nu_0 + i\om_r)e^{-(\nu_0 - i\om_r)|\ta|} - (\nu_0 - i\om_r)e^{-(\nu_0 + i\om_r)|\ta|}]/2i\om_r. \label{4.31} \ee
The terms to be transformed have the same form as the term in Eq. (\ref{2.5}), with the modified damping rates $\nu = \nu_0 - i\om_r$ and $\nu^* = \nu_0 + i\om_r$. Notice that $|\nu|^2 = \om_0^2$. Consider the first contribution to the double integral in Eq. (\ref{1.6}). It follows from Eqs. (\ref{2.10}) and (\ref{4.31}) that this contribution is proportional to
\be \frac{1}{2i\om_r} \biggl[ \frac{\nu^*2\nu T}{\nu^2 + \om^2} - \frac{\nu 2\nu^*T} {(\nu^*)^2 + \om^2} \biggr]
= \frac{2\om_0^2T}{\om_r} \im \biggl[ \frac{1}{\nu^2 + \om^2} \biggr]. \label{4.32} \ee
By multiplying the right side of Eq. (\ref{4.32}) by the factor $\ep^2\si/4\nu_0\om_0^2$, and using the identities $\im[(\nu^*)^2] = 2\nu_0\om_r$ and $|\nu^2 + \om^2|^2 = (\om_0^2 - \om^2)^2 + 4\nu_0^2\om^2$, one obtains the contribution
\be S_1(\om) = \frac{\ep^2\si T}{(\om_0^2 - \om^2)^2 + 4\nu_0^2\om^2}. \label{4.33} \ee
The formula for the second contribution can be determined in the same way. However, it follows from Eq. (\ref{2.10}) that the second contribution is smaller than the first by a factor of order $\nu T$, so it can be omitted. (In this discussion, $\om_0$, $\om_r$ and $\nu_0$ are all of the same order.) 

Now consider the transient terms, which were omitted from correlation (\ref{4.25}).
When one Fourier transforms the transient term in Eq. (\ref{4.22}), the double integral involves exponentials with the arguments
\be -\nu_0(t' + t) + i(\om \pm \om_r)(t' - t) = -[\nu_0 + i(\om \pm \om_r)]t - [\nu_0 - i(\om \pm \om_r)]t'. \ee
Hence, the double integrals ($\pm$) factorize into single integrals and evaluate to $1/[\nu_0^2 + (\om \pm \om_r)^2]$, plus exponentially small terms.
When one transforms the transient term in Eq. (\ref{4.23}), the double integral involves exponentials with the arguments
\be -(\nu_0 \mp i\om_r)(t' + t) + i\om(t' - t) = -(\nu_0 \mp i\om_r + i\om)t - (\nu_0 \mp i \om_r - i\om)t'. \ee
Hence, the double integrals evaluate to $1/[(\nu_0 \mp i\om_r)^2 + \om^2]$, plus exponentially small terms. Thus, the transient contributions to the spectrum are also smaller than contribution (\ref{4.32}) by factors of order $\om_0T$ and can be omitted. The normalized spectral density, which is density (\ref{4.33}) divided by $\ep^2\si T/\om_0^4$, is plotted as a function of frequency in Fig. 5(b). Spectrum (\ref{4.33}) has maxima of $\ep^2\si T/4\nu_0^2\om_r^2$ at $\om = \pm (\om_0^2 - 2\nu_0^2)^{1/2}$ and decreases like $1/\om^4$ as $\om \rightarrow \infty$. 

The frequency-domain versions of Eqs. (\ref{4.1}) and (\ref{4.2}), which are based on the standard derivative rule, can be written in the matrix form
\be \left[\begin{array}{cc} -i\om & -1 \\ \om_0^2 & -i\om + 2\nu_0 \end{array}\right]
\left[\begin{array}{c} X(\om) \\ V(\om) \end{array}\right]
= \left[\begin{array}{c} 0 \\ \ep R(\om) \end{array}\right].\label{4.41} \ee
Equation (\ref{4.41}) has the solution
\be \left[\begin{array}{c} X(\om) \\ V(\om) \end{array}\right] = \frac{1}{\De(\om)}
\left[\begin{array}{c} \ep R(\om) \\ -i\om\ep R(\om) \end{array}\right], \label{4.42} \ee
where the determinant
\be \De(\om) = \om_0^2 - 2i\nu_0\om - \om^2. \label{4.43} \ee
The convolution theorem (App. A) shows that the frequency-domain solutions for $X$ and $V$ are consistent with the time-domain solutions (\ref{4.3}) and (\ref{4.4}). [In App. D, the inverse transform of $1/\De(\om)$ is shown to be $H(t)\sin(\om_rt)e^{-\nu_0t}/\om_r$.]

By combining Eqs. (\ref{3.14}) and (\ref{4.42}), one obtains the spectral energy density
\be \<|X(\om)|^2\> = \frac{\ep^2\si T}{(\om_0^2 - \om^2)^2 + 4\nu_0^2\om^2}. \label{4.45} \ee
One can calculate the spectral energy by doing a contour integral \cite{mck20b}, the result of which is
\be \int_{-\infty}^\infty \<|X(\om)|^2\> d\om/2\pi = \ep^2\si T /4\nu_0\om_0^2. \label{4.46} \ee
Notice that the spectral density is nonsingular, because the measurement time is finite, and the spectral energy is finite, because the damping rate is nonzero. Even weak damping filters the noise source. By comparing Eqs. (\ref{4.45}) and (\ref{4.46}) to Eqs. (\ref{4.33}) and (\ref{4.19}), respectively, one finds that for a weakly-damped oscillator, which attains a stochastic steady state, the standard frequency-domain analysis, which omits $X(T)$, is asymptotically correct. Equations (\ref{4.25}) and (\ref{4.45}) satisfy the Wiener--Khinchin theorem \cite{mck20b}.

Solution (\ref{4.42}) implies that $V(\om) = -i\om X(\om)$. By doing another contour integral \cite{mck20b}, one finds that the spectral energy
\be \int_{-\infty}^\infty \<|V(\om)|^2\> d\om/2\pi = \ep^2\si T /4\nu_0. \label{4.47} \ee
In steady state, the displacement and velocity variances are the associated spectral energies divided by the measurement time. Hence,
\be \<V^2\> = \ep^2\si/4\nu_0 = \om_0^2\<X^2\>. \label{4.48} \ee
The average kinetic and potential (spring) energies are equal.

For a weakly-damped oscillator, most of the spectral energy is contained in the peaks near $\pm \om_0$. Let $\om = \pm \om_0 + \de$, where $\de$, $\nu_0 \ll \om_0$. Then, near each peak, the spectral density
\be S(\de) \approx \ep^2\si T / 4\om_0^2(\nu_0^2 + \de^2). \label{4.51}\ee
The normalized spectral density, which is density (\ref{4.51}) divided by $\ep^2\si T/\om_0^4$, is plotted as a function of frequency in Fig. 6. Spectrum (\ref{4.51}) has a maximum of $\ep^2\si T/4\nu_0^2\om_0^2$ and an effective width of $\pm \nu_0$, and contains the spectral energy $\ep^2\si T/8\nu_0\om_0^2$. [The integral of $1/(\nu_0^2 + \de^2)$ is $\pi/\nu_0$.] The sum of the spectral energies of the two Lorentzian peaks equals the total spectral energy (\ref{4.46}). These results suggest that the strongly-damped-oscillator equation (\ref{3.2}) models the behavior of a resonantly-driven weakly-damped oscillator.
\begin{figure}[h!]
\vspace{-0.1in}
\centerline{\includegraphics[width=2.6in]{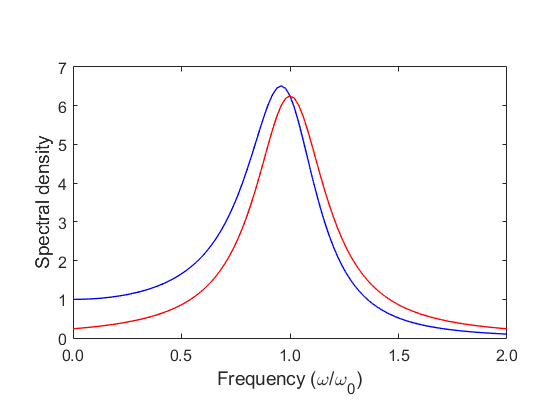}}
\vspace{-0.1in}
\caption{Spectral energy density of a weakly-damped oscillator for $\nu_0/\om_0 = 0.2$. The blue curve represents formula (\ref{4.45}), whereas the red curve represents the Lorentzian formula (\ref{4.51}).}
\end{figure}

It is easy to establish this connection. By combining Eqs. (\ref{4.1}) and (\ref{4.2}), one obtains the second-order equation
\be (d_{tt}^2 + 2\nu_0d_t + \om_0^2)X = \ep R(t). \label{4.52} \ee
Suppose that the displacement
\be X(t) = B(t)e^{-i\om_0t} + B^*(t)e^{i\om_0t}, \label{4.53}\ee
where the amplitude $B$ varies slowly compared to the phase factor. Then, by substituting ansatz (\ref{4.53}) in Eq. (\ref{4.52}), and omitting the terms $d_{tt}^2 B$ and $2\nu_0 d_tB$, one finds that
\be (d_t + \nu_0)B \approx R_b(t), \label{4.54} \ee
where $R_b = i\ep R(t) e^{i\om_0t}/2\om_0$. This source function satisfies an equation similar to (\ref{2.3}), in which the source strength $\si_b = \ep^2\si/4\om_0^2 = \si_v/4\om_0^2$. Thus, a strongly-damped oscillator can be considered as the near-resonance limit of a weakly-damped oscillator.

\section{Undamped oscillator}

Now consider the response of an undamped oscillator (integrator) that is driven by noise. The amplitude equation is
\be d_t A = R_a(t), \label{5.1} \ee
where $R_a$ has the properties (\ref{2.1}) -- (\ref{2.3}).
The solution of Eq. (\ref{5.1}) is
\be A(t) = \int_0^t R(s) ds. \label{5.2} \ee
This solution can be written in the form of Eq. (\ref{3.3}), with $G(t) = H(t)$. Solution (\ref{5.2}) describes a Wiener process, which is also called Brownian motion \cite{gar85,lem02}.
It follows from solution (\ref{5.2}) that the correlation function
\ba \<A^*(t)A(t')\> &= &\int_0^t \int_0^{t'} \<R^*(s)R(s')\> ds'ds \nonumber \\
&= &\si\min(t,t'). \label{5.3} \ea
Noise that is supplied to the oscillator between $t$ and $t'$ affects the later displacement, but not the earlier one, so it cannot affect the correlation.
Correlation (\ref{5.3}) is illustrated in Fig. 7.
Unlike a damped oscillator, at no time does the correlation become a function of the time difference $t' - t$, so the Wiener--Khinchin theorem does not apply to an undamped oscillator. The power $\<|A(t)|^2\>$ increases linearly with time and the energy
\be \int_0^T \<|A(t)|^2\> dt = \si T^2/2 \label{5.4} \ee
is finite, because the measurement time is finite. Notice that the energy increases more rapidly with $T$ for an undamped oscillator than for a damped one [Eq. (\ref{3.7})], because the undamped oscillator integrates the effects of the source.
\begin{figure}[h!]
\vspace{-0.1in}
\centerline{\includegraphics[width=2.8in]{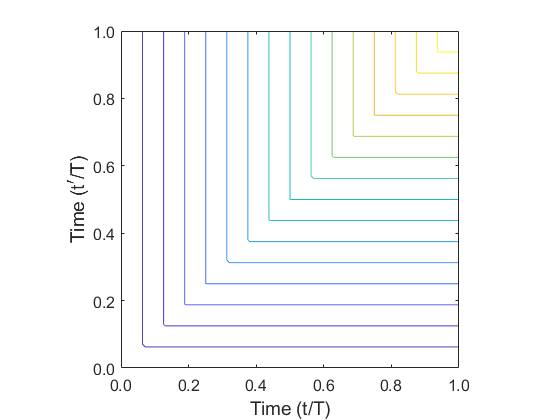}}
\vspace{-0.1in}
\caption{Two-time correlation function of an undamped oscillator. The correlation never becomes a function of the time difference $t' - t$.}
\end{figure}

Despite the fact that the oscillator response is nonstationary, the spectral energy density can be determined. It follows from Eqs. (\ref{1.6}) and (\ref{5.3}) that
\be \<|A(\om)|^2\> = \int_0^T \int_0^T \si\min(t,t') e^{i\om(t' - t)} dtdt'. \label{5.5} \ee
The $t'$-integral in Eq. (\ref{5.5}) is
\ba \int_0^t t' e^{i\om t'} dt' + \int_t^T t e^{i\om t'} dt'
&= &\frac{te^{i\om t}}{i\om} + \frac{e^{i\om t} - 1}{\om^2} + \frac{t(e^{i\om T} - e^{i\om t})}{i\om} \nonumber \\
&= &\frac{e^{i\om t} - 1}{\om^2} + \frac{te^{i\om T}}{i\om} \label{5.6} \ea
and the $t$-integral that follows is
\ba \int_0^T \biggl[ \frac{1 - e^{-i\om t}}{\om^2} + \frac{te^{i\om(T - t)}}{i\om} \biggr] dt
&= &\frac{T}{\om^2} + \frac{e^{-i\om T} - 1}{i\om^3} + \frac{T}{\om^2} + \frac{1 - e^{i\om T}}{i\om^3} \nonumber \\
&= &\frac{2T[1 - \sinc(\om T)]}{\om^2}, \label{5.7} \ea
where $\sinc(x) = \sin(x)/x$. Multiplying this result by $\si$ produces the spectrum
\be \<|A(\om)|^2\> = \frac{2\si T[1 - \sinc(\om T)]}{\om^2}. \label{5.8} \ee
The derivation of Eq. (\ref{5.8}) was based on Eqs. (\ref{1.2}) and (\ref{5.2}). One can check it by using the alternative starting equation
\ba A(\om) &= &\int_0^T R(s) \int_s^T e^{i\om t} dt ds \nonumber \\
&= &\int_0^T \frac{e^{i\om T} - e^{i\om s}}{i\om} R(s) ds, \label{5.9} \ea
which one obtains by changing the order of integration. It follows from Eq. (\ref{5.9}) that
\ba \<|A(\om)|^2\> &= &\int_0^T \int_0^T \frac{e^{-i\om T} - e^{-i\om s}}{-i\om} \frac{e^{i\om T} - e^{i\om s'}}{i\om} \<R^*(s)R(s')\> ds ds' \nonumber \\
&= &\si \int_0^T \frac{e^{-i\om T} - e^{-i\om s}}{-i\om} \frac{e^{i\om T} - e^{i\om s}} {i\om} ds \nonumber \\
&= &\frac{2\si}{\om^2} \int_0^T \{1 - \cos[\om(s - T)] \} ds \nonumber \\
&= &\frac{2\si T[1 - \sinc(\om T)]}{\om^2}, \label{5.10} \ea
as required. The alternative derivation of the spectrum is shorter than the original one.

For low frequencies, $S(\om) \approx \si T^3/3$, which is finite because $T$ is finite, whereas for high frequencies, $S(\om) \approx 2\si T/\om^2$. The spectral energy
\be \int_{-\infty}^\infty |A(\om)|^2 \frac{d\om}{2\pi} = \frac{\si T^2}{\pi} \int_{-\infty}^\infty \frac{1 - \sinc(x)}{x^2} dx. \label{5.11} \ee
By integrating by parts twice, one finds that the $x$-integral is $\pi/2$. Hence, the spectral energy
\be \int_{-\infty}^\infty |A(\om)|^2 d\om/2\pi = \si T^2/2. \label{5.12} \ee
Equation (\ref{5.12}) is consistent with Eq. (\ref{5.4}). Although the oscillator response is nonstationary, it satisfies the Parseval relation (as it should do). It is interesting to note that one can convert Eq. (\ref{5.12}) to Eq. (\ref{3.12}) by replacing one factor of $T$ by $1/\nu$. This result reflects the fact that a detector with measurement time $T$ acts like a frequency filter with bandwidth $\nu = 1/T$.

It is sometimes helpful to replace the complicated function $[1 - \sinc(x)]/x^2$ by the simple Lorentzian $1/(a^2 + x^2)$, which is finite for $x = 0$ and tends to $1/x^2$ as $x \rightarrow \infty$. This Lorentzian has a maximum of $1/a^2$ and an integral of $\pi/a$. By choosing $a = 6^{1/2}$, one reproduces the maximum correctly, but underestimates the integral. In contrast, by choosing $a = 2$, one overestimates the maximum, but reproduces the integral correctly. We prefer the second choice, because the associated spectrum satisfies the Parseval relation. The exact and approximate spectral densities, divided by $2\si T^3$, are plotted as functions of frequency in Fig. 8.
\begin{figure}[h!]
\vspace{-0.1in}
\centerline{\includegraphics[width=2.6in]{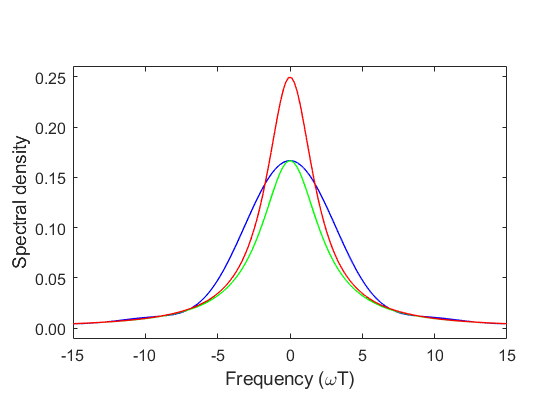}}
\vspace{-0.1in}
\caption{Spectral energy density of an undamped oscillator. The blue curve denotes the exact spectrum (\ref{5.8}), whereas the green and red curves denote Lorentzian approximations with $a = 6^{1/2}$ and 2, respectively.}
\end{figure}

If we had solved the problem in the frequency domain, using the standard derivative rule, we would have found that
\be A(\om) = iR(\om)/\om. \label{5.13} \ee
It follows from this solution that the spectrum
\be \<|A(\om)|^2\> = \si T/\om^2. \label{5.14} \ee
Although these results were easy to obtain, they are wrong in several ways. First, the inverse transform of $i/\om$ is $\sign(t)/2$ (App. D), not $H(t)$, so solution (\ref{5.13}) is not consistent with solution (\ref{5.2}). Second, not only is spectrum (\ref{5.14}) infinite at the origin, its integral is infinite and it underestimates spectrum (\ref{5.8}) by a factor of 2 at high frequencies. These are all serious failures, which were caused by a naive application of the standard derivative rule. The derivative rule is applied carefully in App. E.

When singularities occur in the analysis of a physical system, it is common to add some damping ($\al$) to the system, redo the analysis and take the limit as $\al \rightarrow 0$. By following this prescription, one would obtain Eq. (\ref{5.13}), with $\om$ replaced by $\om + i\al$, and Eq. (\ref{5.14}), with $\om^2$ replaced by $\om^2 + \al^2$. The inverse transform of $i/(\om + i\al)$ is $H(t)e^{-\al t}$ (App. D), which tends to $H(t)$ as $\al \rightarrow 0$. This result is consistent with Eq. (\ref{5.2}). However, the integral of the spectrum is $\si T/2\al$, which tends to $\infty$ as $\al \rightarrow 0$. For an undamped oscillator, the standard prescription does not work.

In this section and the previous ones, only the spectra $S(\om)$ were calculated [Eqs. (\ref{2.4}), (\ref{2.10}), (\ref{3.11}), (\ref{4.33}) and (\ref{5.8})]. The associated frequency correlations $C(\om,\om')$ are calculated in App. F, for completeness.

\section{Discussion}

In Secs. 3 -- 5, the systems (processes) considered were characterized by their means and two-time correlations (from which their variances can be deduced). These quantities provided useful information about the processes, but did not describe them completely. In this section, the probability distributions of the displacements are discussed and the processes are characterized.

Let $X$ denote a random variable (process). Then the probability that $X$ has a value between $x$ and $x + dx$ is $P(x)dx$, where $P$ is the probability distribution function. The process is said to have Gaussian statistics if the probability distribution is the Gaussian function
\be P(x) = \exp[-(x - \mu)^2/2\si^2]/(2\pi\si^2)^{1/2}, \label{6.1} \ee
where $\mu$ and $\si$ are parameters. In the rest of this section, we will use the same notation for $X$ and $x$. The $n$th moment $\<x^n\>$ is defined as the integral
\be \<x^n\> = \int_{-\infty}^\infty x^n P(x) dx. \label{6.2} \ee
For distribution function (\ref{6.1}), the first moment (mean) $\<x\> = \mu$ and the second moment $\<x^2\> = \si^2 + \mu^2$, from which it follows that the variance $\<x^2\> - \<x\>^2 = \si^2$.

The characteristic function
\be C(k) = \<e^{ikx}\> = \int_{-\infty}^\infty P(x)e^{ikx} dx \label{6.3} \ee
is the Fourier transform of the distribution function. It is useful because
\be \<x^n\> = \lim_{k \rightarrow 0} (-i\pd_k)^n C(k) \label{6.4} \ee
and differentiation is usually easier than integration. The distribution function is the inverse transform of the characteristic function, so one can use either function to characterize the process. If the distribution function is Gaussian, then the characteristic function
\be C(k) = \exp(ik\mu - \si^2k^2/2) \label{6.5} \ee
is also Gaussian. Notice that the coefficient of $k^2$ in formula (\ref{6.5}) is proportional to the variance. If $x$ is measured relative to its mean, the $\mu$-terms in Eqs. (\ref{6.1}) and (\ref{6.5}) are absent.

Now let $X = [x_1,x_2 \dots x_n]^t$ denote multiple random variables (displacement and velocity, for example) and $K = [k_1,k_2 \dots k_n]$ denote multiple transform variables. Then this multiple-variable process is said to be Gaussian, with zero means, if its characteristic function
\be C(K) = \exp(-K^tMK/2), \label{6.6} \ee
where $M = [m_{ij}]$ is a matrix. The moments
\ba \<x_1^{n_1}\> &= &\lim_{k_1 \rightarrow 0} (-i\pd_{k_1})^{n_1} C(K), \label{6.7} \\
\<x_1^{n_1}x_2^{n_2}\> &= &\lim_{k_1 \rightarrow 0} \lim_{k_2 \rightarrow 0} (-i\pd_{k_1})^{n_1} (-i\pd_{k_2})^{n_2} C(K), \label{6.8} \ea
and higher moments are defined by similar formulas. By applying formulas (\ref{6.7}) and (\ref{6.8}), 
one finds that the variances $\<x_i^2\> = m_{ii}$ and the correlations $\<x_ix_j\> = m_{ij}$. For this reason, $M$ is called the covariance matrix. By inverse transforming the characteristic function, one obtains the distribution function
\be P(X) = \frac{\exp(-X^tM^{-1}X/2)}{[(2\pi)^n\det(M)]^{1/2}}. \label{6.9} \ee
Notice that the coefficient matrix in formula (\ref{6.9}) is the inverse of the covariance matrix (and {\it vice versa}). Formula (\ref{6.9}) is standard. A convenient derivation of it is provided in App. A of \cite{mck12}.

For many stochastic processes, it is reasonable to assume that the noise source has Gaussian statistics. To be precise, define the (Wiener) increment, or impulse, $W(\ta) = \tint_0^\ta R(s) ds$, where $R(s)$ is the source function and $\ta$ is a short time interval. Then, for white noise, the impulses have a Gaussian distribution function with mean $\<W(\ta)\> = 0$ and variance $\<|W(\ta)|^2\> = \si\ta$, where $\si$ is the source strength. This assumption is common for two complementary reasons.
First, suppose that $\hat{a}$ is the amplitude operator of a quantum mode, in which case $\<\hat{a}^\d\hat{a}\>$ is the expected number of quanta. Then, if the mode is in a coherent state, the expected quadratures have Gaussian probability distributions with variances of 1/2 \cite{lou64,mck12}. If the mode experiences linear amplification or attenuation, the quadrature means and variances change, but their distributions remain Gaussian. One can reproduce this behavior semi-classically by adding noise terms with Gaussian statistics to the classical amplitude equation. (Gain and loss both degrade the signal-to-noise ratio.)
Second, if a classical diplacement is driven by many independent impulses, the central limit theorem \cite{gar85,lem02} ensures that the displacement statistics are Gaussian, even if the impulse statistics are not.

A time-dependent random process is said to be strictly stationary if the $n$th moment $\<x(t_1)x(t_2) \dots x(t_n)\>$ is independent of the origin of time. In other words, one can replace $t_1$, $t_2$, $\dots$ $t_n$ by $t_1 - \ta$, $t_2 - \ta$, $\dots$ $t_n - \ta$ without changing the moment. A process is said to be weakly (or wide-sense) stationary if the first two moments $\<x(t_1)\>$ and $\<x(t_1)x(t_2)\>$ are independent of the time origin. For such a process, the mean is constant and the two-time correlation is a function of only the time difference $t_2 - t_1$. Because the mean is constant, one can simplify the analysis of the process by measuring $x(t)$ relative to the mean.

A time-dependent process is said to have Gaussian statistics if its $n$-sample characteristic function has the form (\ref{6.6}). If such a process is weakly stationary, it is also strictly stationary, because the correlation condition applies to arbitrary $t_1$ and $t_2$ (or $t_i$ and $t_j$). Hence, in the covariance matrix, each entry $m_{ij}$ is a function of $t_j - t_i$, which is independent of the time origin. Consequently, so also is the $n$th moment, which is determined completely by the covariance matrix.

In Secs. 3 -- 5, the oscillator displacements (\ref{3.5}), (\ref{4.3}) and (\ref{5.2}) are sums of Gaussian impulses multipled by deterministic response (Green) functions, so the displacements also have Gaussian statistics, which are specified completely by their means (of zero), and correlations (\ref{3.10}), (\ref{4.25}) and (\ref{5.3}), from which the variances follow. In steady state, strongly- and weakly-damped oscillation are weakly-stationary processes. Hence, as explained above, they are also stationary. In contrast, undamped oscillation is a nonstationary process, because the variance of the displacement increases linearly with time. Nonetheless, undamped oscillation has stationary increments: The displacement difference
\be \de X(t,\ta) = X(t + \ta) - X(t) = \int_t^{t+\ta} R(s) ds. \label{6.10} \ee
Because the source term in Eq. (\ref{6.10}) is stationary, the integral (increment) does not depend on the time origin $t$, so the displacement difference is a function of $\ta$ only: $\de X(\ta) = \tint_0^\ta R(s) ds$. Thus, the analyses of Secs. 3 -- 5 provide examples of the three main classes of stationarity, about which much is known \cite{pap91}.

\section{Summary}

In this tutorial, three examples of stochastic systems were discussed: A strongly-damped oscillator, a weakly-damped oscillator and an undamped oscillator (integrator) driven by noise. The dynamics of these systems are characterized by the temporal correlation functions and spectral densities of their displacements [$\<A^*(t)A(t')\>$ and $\<|A(\om)|^2\>$, respectively], which were determined and discussed.

After transients associated with the initial conditions have died out, strongly- and weakly-damped oscillators attain stochastic steady states, in which the temporal correlations (\ref{3.10}) and (\ref{4.25}) depend on the time difference $|t - t'|$, rather than $t$ and $t'$ separately. The correlations are decreasing functions of the time difference, because the effects of noise impulses only influence the displacements for times of the order of the damping time. (These systems have finite memory.) Because these systems (processes) are weakly stationary, the Wiener--Khinchin theorem applies and their spectral energy densities (\ref{3.11}) and (\ref{4.33}) are the Fourier transforms of their temporal correlations with respect to the time difference, multiplied by the measurement time $T$. Although the idealized noise model on which these results are based (white noise) has infinite energy, the spectral energy densities and integrated spectral energies are finite, because damping filters the applied noise.

In contrast, an undamped oscillator never reaches a steady state. The oscillator power increases monotonically with time, because the oscillator integrates the effects of the applied noise impulses. (This system has infinite memory.) The correlation (\ref{5.3}) is a function of $\min(t,t')$, because the noise applied between $t$ and $t'$ has no effect on the correlation of the earlier and later displacements. The spectrum (\ref{5.8}) has a sharp peak at zero frequency, but the spectral energy density and integrated spectral energy are finite if the measurement time is finite. Because this system is nonstationary, the Wiener--Khinchin theorem does not apply.

For each system, the dynamics were analyzed in the time domain (by solving differential equations) and in the frequency domain (by solving algebraic equations). The frequency-domain analyses were based on the standard derivative rule that the Fourier transform of $d_t A$ is the transform of $A$ multiplied by $-i\om$. For damped oscillators, the frequency-domain analyses miss the transient displacements, but predict the time-asymptotic steady states accurately. (The relative errors in the spectral densities and energies are of order $1/\nu T$, where $\nu$ is the damping rate.) However, for an undamped oscillator, the frequency-domain analysis produces results that are incorrect and singular. These failures are caused by the omission of the final value $A(T)$ in the derivative rule. When one models driven systems, in which the displacements do not tend to zero as time tends to infinity, one should use the derivative rule with caution.

The mathematical methods required to study randomly-driven oscillators and the physical insights one gains from such a study can be applied to a variety of problems, in fields such as biology, chemistry, finance and optimization \cite{ber83,kam07,bax96,mar15}. In a subsequent tutorial \cite{mck21}, we will use the methods and insights discussed herein to study laser linewidths (phase fluctuations).
This tutorial was restricted to linear oscillators subject to additive noise. A review of linear and nonlinear oscillators, subject to additive and multiplicative noise, can be found in \cite{git05}.

\section*{Acknowledgement}

We thank the reviewers for bringing to our attention \cite{ber83,kam07,mar15,git05}.

\section*{Data availability}

No data were generated or analyzed in this research.

\section*{Disclosure}

The authors declare that there is no conflict of interest.

\section*{Funding}

TJS and ASH were supported by the Natural Sciences and Engineering Research Council of Canada.

\newpage

\section*{Appendix A: Fourier transforms}

Let $A_2(t)$ be an integrable two-sided function of time (defined for positive and negative times). Then the forward and backward Fourier transforms are defined by the equations
\ba A_2(\om) &= &\int_{-\infty}^\infty A_2(t)e^{i\om t} dt, \label{a1} \\
A_2(t) &= &\int_{-\infty}^\infty A_2(\om)e^{-i\om t} d\om/2\pi. \label{a2} \ea
In both integrals, the independent variables are real and range from $-\infty$ to $\infty$.
Now let $B_2(t)$ be another integrable two-sided function. Then the mathematical cross-correlation of $A_2$ and $B_2$ is defined by the integral
\be C_2(\ta) = \int_{-\infty}^\infty A_2^*(t)B_2(t + \ta) dt, \label{a3} \ee
where the time difference $\ta$ is arbitrary. (The physical correlation involves integration from 0 to $T$ and division by $T$, so that the correlation has the same units as the product of the amplitudes.) Let $C_2(\om)$ be the Fourier transform of the auto-correlation with respect to the time difference. Then the correlation theorem states that
\be C_2(\om) = A_2^*(\om)B_2(\om), \label{a4} \ee
where $A_2^*(\om)$ is an abbreviation for $[A_2(\om)]^*$. For the special case in which the functions are identical, $C_2(\om) = |A_2(\om)|^2$. The Fourier transform of the temporal auto-correlation is the spectral energy density. Likewise, the inverse transform of the spectrum is the auto-correlation. For the special case in which $\ta = 0$, the correlation theorem reduces to the Parseval equation
\be \int_{-\infty}^\infty |A_2(t)|^2 dt = \int_{-\infty}^\infty |A_2(\om)|^2 d\om/2\pi,  \label{a5} \ee
which states that the time- and frequency-domain definitions of energy are consistent.

The convolution is defined by the integral
\be D_2(\ta) = \int_{-\infty}^\infty A_2(t)B_2(\ta - t) dt, \label{a6} \ee
where $\ta$ is arbitrary. In contrast to the correlation integral, $A_2$ is not conjugated and $B_2$ is a function of $\ta - t$, not $\ta + t$. The convolution theorem states that
\be D_2(\om) = A_2(\om)B_2(\om). \label{a7} \ee
It is used in analyses of linear systems, in which context $A_2$ is the driving term and $B_2$ is the response (Green) function.

This tutorial involves the solutions of initial-value problems, which are defined for non-negative times. If these solutions were integrable, one could derive one-sided versions of the correlation and convolution theorems from the two-sided versions by replacing $A_2(t)$ with $A_1(t) = A_2(t)H(t)$, where $H$ is the Heaviside step function. However, in the problems considered, the dependent variables are driven by noise, which is always present, so the solutions are not integrable. One can prevent singularities in the analyses of these problems by limiting the integration (measurement) time to the large, but finite, value $T$, and one can derive finite-time versions of the correlation and convolution theorems from the one-sided versions by replacing $A_1(t)$ with $A_T(t) = A_1(t)H(T - t) = A_2(t)H(t)H(T - t)$. In this appendix, finite-time results are derived directly from the two-sided results stated above. One can deduce the corresponding one-sided results by letting $T \rightarrow \infty$ and assuming integrability.

The finite-time amplitudes $A_T(t)$ and $B_T(t)$ are only nonzero for $0 \le t \le T$, so the Fourier transform
\be A_T(\om) = \int_0^T A_T(t) e^{i\om t} dt \label{a11} \ee
and the correlation
\be C_T(\ta) = \int_0^T A_T^*(t)B_T(t + \ta) dt. \label{a12} \ee
On the right side of Eq. (\ref{a12}), the second function is nonzero for $-t \le \ta \le T - t$. Because the extreme values of $t$ are 0 and $T$, the extreme values of $\ta$ are $-T$ and $T$. Hence, the Fourier transform
\be C_T(\om) = \int_{-T}^T C_T(\ta) e^{i\om\ta} d\ta. \label{a13} \ee
The integration region for the correlation theorem is illustrated in Fig. 9(a).

By combining Eqs. (\ref{a12}) and (\ref{a13}), and reversing the order of integration, one finds that
\ba C_T(\om) &= &\int_0^T \int_{-T}^T A_T^*(t) e^{-i\om t} B_T(t + \ta) e^{i\om(t + \ta)} d\ta dt, \nonumber \\
&= &\int_0^T A_T^*(t) e^{-i\om t} \int_{-t}^{T-t} B_T(t + \ta) e^{i\om(t + \ta)} d\ta dt \nonumber \\
&= &\int_0^T A_T^*(t) e^{-i\om t} dt \int_0^T B_T(t') e^{i\om t'} dt' \nonumber \\
&= &A_T^*(\om)B_T(\om). \label{a14} \ea
The second of Eqs. (\ref{a14}) was obtained from the first by using the step functions that are implicit in $B_T$.
It is also instructive to consider the inverse transform
\ba C_T(\ta) &= &\int_{-\infty}^\infty \int_0^T \int_0^T A_T^*(t) B_T(t') e^{i\om(t' - t - \ta)} dt dt' d\om/2\pi \nonumber \\
&= &\int_0^T \int_0^T A_T^*(t) B_T(t') \de(t' - t - \ta) dt dt' \nonumber \\
&= &\int_0^T A_T^*(t)B_T(t + \ta) dt. \label{a15} \ea
The second of Eqs. (\ref{a15}) shows that $C_T(\ta)$ is defined for $-T \le \ta \le T$, as stated above.
\begin{figure}[h!]
\vspace*{-0.0in}
\centerline{\includegraphics[height=2.0in]{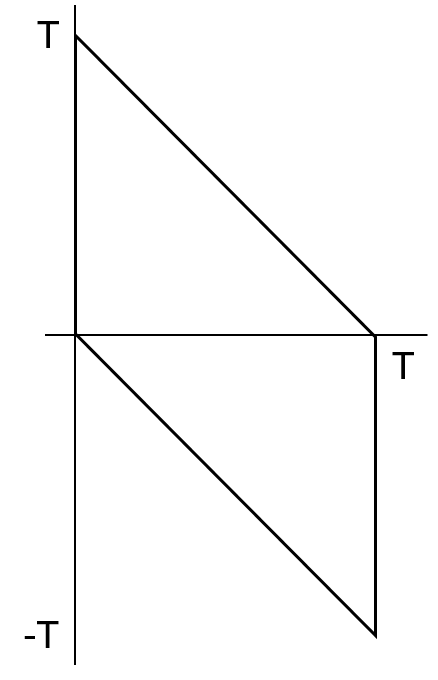} \hspace{0.3in} \includegraphics[height=2.1in]{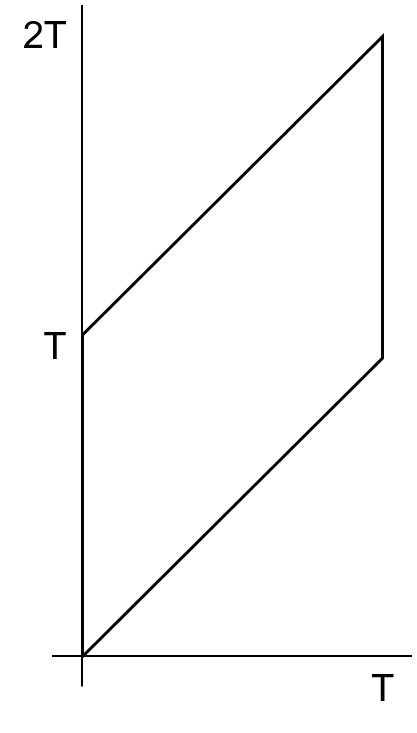}}
\vspace*{-0.0in}
\caption{Integration regions in the $t$-$\ta$ plane for the correlation theorem (left) and the convolution theorem (right). The horizontal axis represents $t$ and the vertical axis represents $\ta$.}
\end{figure}

The finite-time convolution
\be D_T(\ta) = \int_0^T A_T(t)B_T(\ta - t) dt. \label{a16} \ee
On the right side of Eq. (\ref{a16}), the second function is nonzero for $t \le \ta \le T + t$, so the formal upper limit $T$ can be replaced by the effective limit $\ta$. Because the extreme values of $t$ are 0 and $T$, the extreme values of $\ta$ are 0 and $2T$. Hence, the Fourier transform
\be D_T(\om) = \int_0^{2T} D_T(\ta) e^{i\om\ta} d\ta. \label{a17} \ee
The integration region for the convolution theorem is illustrated in Fig. 9(b).

By combining Eqs. (\ref{a16}) and (\ref{a17}), and reversing the order of integration, one finds that
\ba D_T(\om) &= &\int_0^T \int_0^{2T} A_T(t) e^{i\om t} B_T(\ta - t) e^{i\om(\ta - t)} d\ta dt, \nonumber \\
&= &\int_0^T A_T(t) e^{i\om t} \int_{t}^{T+t} B_T(\ta - t) e^{i\om(\ta - t)} d\ta dt \nonumber \\
&= &\int_0^T A_T(t) e^{i\om t} dt \int_0^T B_T(t') e^{i\om t'} dt' \nonumber \\
&= &A_T(\om)B_T(\om). \label{a18} \ea
The second of Eqs. (\ref{a18}) was obtained from the first by using the step functions that are implicit in $B_T$.
One also finds that the inverse transform
\ba D_T(\ta) &= &\int_{-\infty}^\infty \int_0^T \int_0^T A_T(t) B_T(t') e^{i\om(t' + t - \ta)} dt dt' d\om/2\pi \nonumber \\
&= &\int_0^T \int_0^T A_T(t) B_T(t') \de(t' + t - \ta) dt dt' \nonumber \\
&= &\int_0^T A_T(t)B_T(\ta - t) dt. \label{a19} \ea
The second of Eqs. (\ref{a19}) shows that $D_T(\ta)$ is defined for $0 \le \ta \le 2T$, as stated above.
Thus, the correlation and convolution theorems for finite integration times are similar to those for infinite times. The only differences are the limits of integration in definitions (\ref{a13}) and (\ref{a17}).

The preceding discussion of the convolution theorem was based on the assumption that the functions $A$ and $B$ are both defined for arguments between 0 and $T$. Solutions (\ref{3.3}), (\ref{4.3}) and (\ref{5.2}) were written in the Green-function form
\be A(t) = \int_0^t G(t - s)R(s) ds, \label{a21} \ee
where $0 \le s \le t$ (causality) and $t \le T$ (finite time). The transformed amplitude
\ba A_T(\om) &= &\int_0^T \int_0^t G(t - s)R(s) e^{i\om t} ds dt \nonumber \\
&= &\int_0^T \int_s^T G(t - s)R(s) e^{i\om(t-s+s)} dt ds \nonumber \\
&= &\int_0^T R(s) e^{i\om s} \int_0^{T-s} G(\ta) e^{i\om\ta} d\ta ds, \label{a22} \ea
where $\ta = t - s$. The right side of Eq. (\ref{a22}) is not a product of two independent finite-time transforms.

For a strongly-damped oscillator, the transformed Green function $G_T(\om) = 1/(\nu - i\om)$, and for a weakly-damped oscillator, $G_T(\om) = \ep/(\om_0^2 - 2i\nu_0\om - \om^2)$, where exponentially small terms were omitted. By squaring these functions and multiplying by the noise spectral density $\si T$, one obtains the largest contribution to spectrum (\ref{3.11}) and spectrum (\ref{4.33}), respectively. Thus, for damped oscillators, formula (\ref{a18}) produces spectra that are asymptotically correct.

For an undamped oscillator, $G(t) = H(t)$, so its transform
\be G_T(\om) = (e^{i\om T} - 1)/i\om \label{c8} \ee
and the associated spectrum
\be \<|A_T(\om)|^2\> = 2\si T[1 - \cos(\om T)]/\om^2, \label{c9} \ee
where nothing was omitted. Formula (\ref{c9}) is much better than the naive formula $\si T/\om^2$. It has a maximum of $\si T^3$ at $\om = 0$, an integral of $2\pi\si T^2$ and is proportional to $2\si T/\om^2$ (with the correct factor of 2). However, it differs from the exact formula (\ref{5.8}), which has a maximum of $\si T^3/3$ and an integral of $\pi\si T^2$. Thus, for the systems of interest to us, the convolution theorem does not apply as written, and for an undamped oscillator, the relative error is of order 1, which does not tend to zero as $T$ tends to infinity. It is for this reason that we did not use the standard formula $A(\om) = G(\om)R(\om)$ to calculate spectra.

\section*{Appendix B: Double Fourier transforms}

Equation (\ref{1.6}) shows that one can always determine the spectral energy density of a stochastic process from the two-time correlation. However, one cannot always determine the temporal correlation from the spectrum, because there is only one frequency variable ($\om)$, but two time variables ($t$ and $t'$) are required. (If one already knows that the process is stationary, then one can apply the Wiener--Khinchin theorem, which is stated in App. C.) In this appendix, the relations between the time- and frequency-domain characterizations of stochastic processes are discussed.

Define the two-time correlation $C(t,t') = \<A^*(t)A(t')\>$ and the two-frequency correlation $C(\om,\om') = \<A^*(\om)A(\om')\>$, and suppose that both functions are integrable. Then they are linked by the double Fourier transforms
\ba C(\om,\om') &= &\int_{-\infty}^\infty \int_{-\infty}^\infty C(t,t') e^{i\om't' - i\om t} dtdt', \label{b1} \\
C(t,t') &= &\int_{-\infty}^\infty \int_{-\infty}^\infty C(\om,\om') e^{i\om t - i\om't'} d\om d\om'/(2\pi)^2. \label{b2} \ea
For reference, recall that
\be \int_{-\infty}^\infty e^{\pm i\om t} d\om/2\pi = \de(t), \label{b3} \ee
from which it follows that
\be \int_{-\infty}^\infty e^{\pm i\om t} dt = 2\pi\de(\om). \label{b4} \ee

Now consider some examples. Suppose that the correlation,
\be C(t,t') = S(t)\de(t - t'), \label{b5} \ee
is $\de$-correlated in time. Then its transform,
\be C(\om,\om') = S(\om' - \om), \label{b6} \ee
is a function of the frequency difference $\om' - \om$. For the special case in which $S(t) = \si$, where $\si$ is constant, $C(\om,\om') = 2\pi\si\de(\om - \om')$.
Conversely, if the correlation,
\be C(\om,\om') = 2\pi S(\om)\de(\om - \om'), \label{b7} \ee
is $\de$-correlated in frequency, then its inverse transform,
\be C(t,t') = S(t' - t), \label{b8} \ee
is a function of the time difference $t' - t$. For the special case in which $S(\om) = \si$, then $C(t,t') = \si\de(t - t')$, as stated above. If we had used frequency $f$ rather than angular frequency $\om$, the factors of $2\pi$ in the preceding results would have been absent, because $a\de(ax) = \de(x)$.
The bidirectional relations described by Eqs. (\ref{b5}) -- (\ref{b8}) are well known. However, they are also problematic. In the first relation, the power $\<|A(t)|^2\> = S(t)\de(0)$ is undefined, whereas in the second relation, the spectrum $\<|A(\om)|^2\> = 2\pi S(\om)\de(0)$ is undefined.

Because the displacements and correlation functions associated with driven processes are not integrable, one replaces infinite time integrals by finite ones. In the one-sided formalism, the amplitude $A_T(t)$ is nonzero for $0 \le t \le T$. The frequency correlation
\be C_T(\om,\om') = \int_0^T \int_0^T C_T(t,t') e^{i\om't' - i\om t} dtdt', \label{b21} \ee
and the time-integral identity
\be \int_0^T e^{\pm i\om t} dt = (e^{i\om T} - 1)/i\om = 2\pi \de_T(\om), \label{b22} \ee
where $\de_T$ is the complex $\de$-function that is discussed in App. D (see Fig. 11). Notice that $2\pi\de_T(0) = T$.

Now reconsider the previous examples with finite time integrals. If the temporal correlation $C_T(t,t') = S_T(t)\de(t - t')$, then its transform
\ba C_T(\om,\om') &= &\int_0^T S_T(t) e^{i(\om'-\om)t} dt = S_T(\om' - \om). \label{b23} \ea
For the special case in which $S_T(t) = \si$, $C_T(\om,\om') = 2\pi\si\de_T(\om - \om')$, where the effective $\de$-function was defined in Eq. (\ref{b22}). The associated spectrum $\<|A_T(\om)|^2\> = \si T$.
If the correlation $C_T(t,t') = S_T(t' - t)$, where $-T \le t' - t \le T$, then its transform
\ba C_T(\om,\om') &= &\int_0^T \int_0^T S_T(t' - t)e^{i\om'(t'-t)} e^{i(\om' - \om)t} dt dt' \nonumber \\
&= &\int_0^T e^{i(\om' - \om)t} \int_{-t}^{T-t} S_T(\ta) e^{i\om'\ta} d\ta dt \nonumber \\
&= &\int_{-T}^T S_T(\ta) e^{i\om'\ta} \int_{\max(0,-\ta)}^{\min(T-\ta,T)} e^{i(\om'-\om)t} dt d\ta \nonumber \\
&\sim &2\pi S_T(\om') \de_T(\om' - \om). \label{b24} \ea
The integration region in the $t-\ta$ plane was illustrated in Fig. 9(a). A related integral is considered carefully in App. C (see Fig. 10). For now, it is sufficient to know that if $T$ is large and $S_T(\ta)$ is a decreasing function of $|\ta|$, then Eq. (\ref{b24}) is asymptotically correct.
Equations (\ref{b22}) -- (\ref{b24}) are equivalent to Eqs. (\ref{b4}) -- (\ref{b8}), including the factors of $2\pi$. The only difference between the two sets of equations is that the the standard $\de$-function, which is singular, is replaced by the effective $\de$-function, which is nonsingular. Using finite time intervals removes the singularities in the spectra.

\section*{Appendix C: Wiener--Khinchin theorem}

In Sec. 2, the spectral power densities of white and colored noise were shown to be the Fourier transforms of their temporal correlation functions with respect to the time difference. In Secs. 3 and 4, this relation, which is called the Wiener--Khinchin theorem, was demonstrated for strongly-and weakly-damped oscillators that have attained stochastic steady states. This theorem is similar to the auto-correlation theorem of App. A. The difference is that the correlation theorem involves the time integral of an integrable function, whereas the Wiener--Khinchin theorem involves the ensemble average of a nonintegrable function.

The inverse transform of the spectral energy density is
\ba \int_{-\infty}^\infty \<|A_T(\om)|^2\> e^{-i\om\ta} d\om/2\pi
&= &\int_{-\infty}^\infty \int_0^T  \int_0^T \<A_T^*(t)A_T(t')\> e^{i\om(t'-t-\ta)} dtdt'd\om/2\pi \nonumber \\
&= &\int_0^T \int_0^T \<A_T^*(t)A_T(t')\> \de(t'-t-\ta) dtdt' \nonumber \\
&= &\int_{\max(0,-\ta)}^{\min(T-\ta,T)} \<A_T^*(t)A_T(t + \ta)\> dt, \label{c2} \ea
where $-T \le \ta \le T$. For a damped system, which attains a weakly-stationary state, the correlation depends on only the time difference [$C(t,t + \ta) = C(0,\ta) = C(\ta)$]. For such a system, the correlation is constant along horizontal lines in Fig. 10. It follows from the Cauchy--Schwartz inequality that $|C(\ta)| < C(0)$, so the largest contribution to the integral in Eq. (\ref{c2}) comes from the line $\ta = 0$. For a damped system, the correlation is a decreasing function of $|\ta|$, with decay rate $\nu$. Only the lines with $|\ta| < 1/\nu$ contribute significantly to the integral. The contributing region is almost a rectangle of width $T$ and height $2/\nu$, so
\be S(\om) \sim T\int_{-1/\nu}^{1/\nu} C(\ta) e^{i\om\ta} d\ta. \label{c3} \ee
The integration error associated with the missing triangles is negative, and is of relative order $1/\nu T$. Because regions with $|\ta| > 1/\nu$ do not contribute significantly to the integral, one can extend the limits of integration and rewrite Eq. (\ref{c3}) in the standard form
\be S(\om)/T \sim \int_{-T}^T C(\ta) e^{i\om\ta} d\ta. \label{c4} \ee
Thus, the spectral power density is asymptotic to the Fourier transform of the temporal correlation function with respect to the time difference.
\begin{figure}[h]
\vspace*{-0.0in}
\centerline{\includegraphics[width=1.8in]{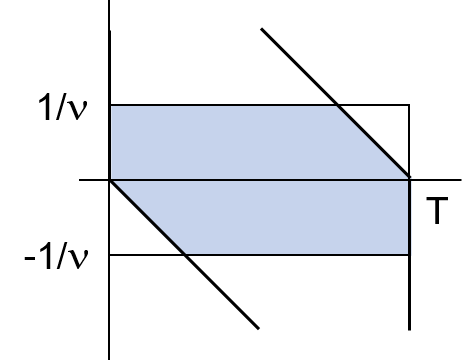}}
\vspace*{-0.0in}
\caption{Integration region in the $t$-$\ta$ plane for the Wiener--Khinchin theorem. The horizontal axis represents $t$ and the vertical axis represents $\ta$. The shaded region contributes to the integral.}
\end{figure}

For white noise, only the line $\ta = 0$ contributes to the integral in Eq. (\ref{c2}), so Eq. (\ref{c4}) is exact. For colored noise, Eq. (\ref{c4}) is asymptotic, as written. For a damped system that starts from rest, the values of the correlation $C(t,t + \ta)$ are smaller than the steady-state value $C(\ta)$ for $0 < t < 1/\nu$ and comparable to it for $1/\nu < t < T$. The lengths of the left line segments (one line for each value of $\ta$) are much shorter than those of the right segments, so Eq. (\ref{c4}) also applies when the transient response is included.

\section*{Appendix D: Specific Fourier transforms}

In this appendix, the forward and backward Fourier transforms of some functions that appear in main text are calculated. For a strongly-damped oscillator (Sec. 3), the time-domain Green function
\be G(t) = H(t)e^{-\nu t}, \label{d1} \ee
where the Heaviside step function
\be H(t) = \left\{ \begin{array}{ccc} 0 & {\rm if} &t< 0, \\
1/2 & {\rm if} &t = 0, \\ 1 & {\rm if} &t > 0. \end{array} \right. \label{d2} \ee
Notice that this Green function is integrable. It follows from Eqs. (\ref{a1}) and (\ref{d1}) that the frequency-domain Green function
\be G(\om) = 1/(\nu - i\om) = (\nu + i\om)/(\nu^2 + \om^2). \label{d3} \ee

If one uses contour integration to evaluate the inversion integral (\ref{a2}), one needs to complete the contour in the upper (lower) half plane for $t < 0$ ($t > 0$).
The Green function in Eq. (\ref{d2}) is analytic in the upper half-plane, but has a pole in the lower half-plane at $\om = -i\nu$. Hence, for $t < 0$, $G(t) = 0$. For $t = 0$, the imaginary part of the inversion integral is 0, whereas the real part is 1/2 [because the integral of $1/(\nu^2 + \om^2)$ is $\pi/\nu$]. For $t > 0$, the inversion integral is $-2\pi i/2\pi$ multiplied by the residue $i e^{-\nu t}$, which is $e^{-\nu t}$. Hence, the inverse transform of the frequency-domain Green function equals the time-domain Green function (as it should do).

For an undamped oscillator (Sec. 5), the time-domain Green function
\be G(t) = H(t). \label{d11} \ee
Unfortunately, this Green function is not integrable. The related function $H(t)e^{-\al t}$ is integrable when $\al$ is positive, so one can  try to fix the non-integrability problem by using the limit function
\be G_\al(t) = \lim_{\al \rightarrow 0} H(t) e^{-\al t} \label{d12}. \ee
It follows from Eqs. (\ref{d3}) and (\ref{d12}) that
\be G_\al(\om) = \lim_{\al \rightarrow 0} (\al + i\om)/(\al^2 + \om^2). \label{d13} \ee
How should one interpret this limit function? The real part has a peak at $\om = 0$, of height $1/\al$ and width $\al$, decreases like $1/\om^2$ as $\om \rightarrow \infty$ and its integral is $\pi$. The imaginary part is an odd function of $\om$, decreases like $1/\om$ as $\om \rightarrow \infty$ and its integral is 0. Hence, as $\al \rightarrow 0$,
\be G_\al(\om) \rightarrow \pi\de(\om) + i/\om. \label{d14} \ee
Formula (\ref{d14}) is standard \cite{bra00}. It is illustrated in Fig. 11.

The backward transform of the first term on the right side of Eq. (\ref{d14}) is 1/2, for all times. The inversion integral for the second term is
\be \int_{-\infty}^\infty [i\cos(\om t) + \sin(\om t)] d\om / 2\pi\om. \label{d15} \ee
The real part of this integral is $\sign(t)/2$ [because the integral of $\sinc(x)$ is $\pi$ and sine is an odd function of $t$], whereas the imaginary part has a principal value of 0. Thus, the inverse transform of the frequency-domain limit function equals the time-domain limit function $H(t)$. The limit procedure used to obtain Eq. (\ref{d14}) is sensible. 

Another important function (which was just mentioned) is the sign function
\be \Si(t) = \left\{ \begin{array}{ccc} -1 & {\rm if} &t< 0, \\
0 & {\rm if} &t = 0, \\ 1 & {\rm if} &t > 0. \end{array} \right. \label{d16} \ee
Notice that $\Si(t) = H(t) - H(-t)$ is the odd extension of the step function (and 1 is the even extension). This function is also not integrable, so one defines the limit function
\be \Si_\al(t) = \lim_{\al \rightarrow 0} \Si(t)e^{-\al|t|}. \label{d17} \ee
It follows from Eqs. (\ref{d3}) and (\ref{d17}) that
\ba \Si_\al(\om) &= &\lim_{\al \rightarrow 0} [1/(\al - i\om) - 1/(\al + i\om)] \nonumber \\
&= &\lim_{\al \rightarrow 0} 2i\om/(\al^2 + \om^2). \label{d18} \ea
The function on the right side of Eq. (\ref{d18}) is 0 at $\om = 0$, has large extremes at $\om = \pm\al$ and decreases like $2i/\om$ as $\al \rightarrow 0$. Hence, as $\al \rightarrow 0$,
\be \Si_\al(\om) \rightarrow 2i/\om. \label{d19} \ee
Formula (\ref{d19}) is also standard \cite{bra00}. It is illustrated (divided by 2) in Fig. 11(b) and its inverse transform was discussed after Eq. (\ref{d15}).

\begin{figure}[h]
\vspace*{-0.1in}
\centerline{\includegraphics[width=2.6in]{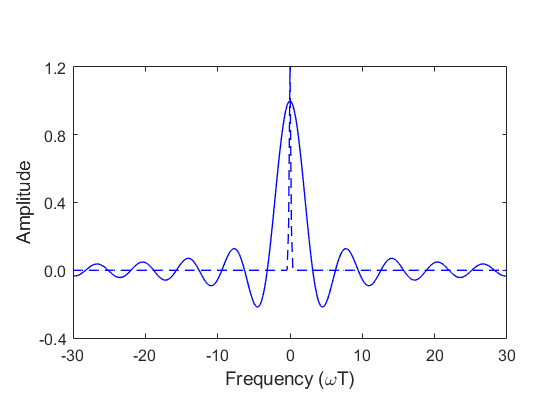} \hspace{0.0in} \includegraphics[width=2.6in]{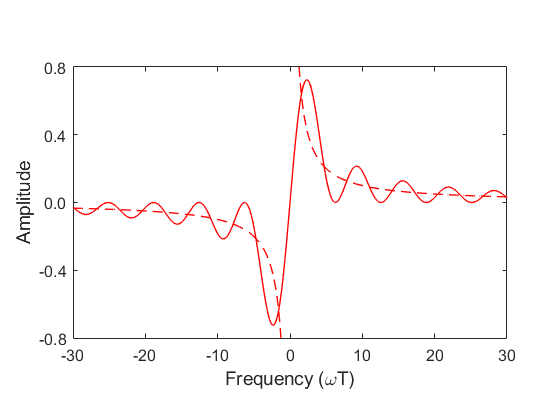}}
\vspace*{-0.1in}
\caption{Real (left) and imaginary (right) parts of the Fourier transform of the step function. The dashed curves represent formula (\ref{d14}), whereas the solid curves represent formula (\ref{d21}).}
\end{figure}

One can also try to fix the non-integrability problems by using finite-time transforms, in which context the truncated step function $H_T(t) = H(t)H(T - t)$. Its transform
\ba H_T(\om) &= &(e^{i\om T} - 1)/i\om \nonumber \\
&= &\sin(\om T)/\om + i[1 - \cos(\om T)]/\om. \label{d21} \ea
This transform (divided by $T$) is illustrated in Fig. 11. The real part of the transform has a peak at $\om = 0$, of height $T$ and width $\pi/T$, and its integral is $\pi$. The imaginary part is an odd function of $\om$, its local average decreases like $1/\om$ as $\om \rightarrow \infty$ and its integral is 0. Hence, for large values of $T$,
\be H_T(\om) \approx \pi\de(\om) + i/\om. \label{d22} \ee
Formula (\ref{d22}) is consistent with formula (\ref{d14}).

Likewise, the truncated sign function $\Si_T(t) = \Si(t)H(t + T)H(T - t)$ and its transform
\ba \Si_T(\om) &= &(e^{i\om T} - 1)/i\om  - (e^{-i\om T} - 1)/(-i\om) \nonumber \\
&= &2i[1 - \cos(\om T)]/\om. \label{d23} \ea
This transform (divided by $2T$) is illustrated in Fig. 11(b). The function on the right side of Eq. (\ref{d23}) was discussed after Eq. (\ref{d21}). For large values of $T$,
\be \Si_T(\om) \approx 2i/\om. \label{d24} \ee
Formula (\ref{d24}) is consistent with formula (\ref{d19}).

Now consider the inverse transforms of formulas (\ref{d21}) and (\ref{d23}). If the time-integration domain is extended slightly to the left of $-T$ and slightly to the right of $T$, the end values of the step function are 0, in which case one can apply the rule that the transform of $d_t A(t)$ is the transform of $A(t)$ multiplied by $-i\om$. Conversely, the inverse transform of $iA(\om)/\om$ is the integral $\tint_{-T}^t A(s) ds$. Hence, the inverse transform of $H_T(\om)$ is
\ba &&\int_{-T}^t \int_{-\infty}^\infty (1 - e^{i\om T}) e^{-i\om s} d\om ds/2\pi \nonumber \\
&= &\int_{-T}^t [\de(s) - \de(s - T)] ds \ = \ H_T(t). \ea
Likewise, the inverse transform of $\Si_T(\om)$ is
\ba &&\int_{-T}^t \int_{-\infty}^\infty (2 - e^{i\om T} - e^{-i\om T}) e^{-i\om s} d\om ds/2\pi \nonumber \\
&= &\int_{-T}^t [-\de(s + T) + 2\de(s) - \de(s - T)] ds \ = \ \Si_T(t). \ea
Thus, in the finite-time approach, the inverse transforms of the frequency-domain functions equal the original time-domain functions (as they should do).

Suppose that $F(t)$ describes a causal response, which equals 0 for $t < 0$. Then its Fourier transform $F(\om)$ is analytic in the upper half-plane and tends to 0 as $|\om| \rightarrow \infty$. By integrating along a contour that consists of the real $\om'$-axis, a small semicircle around the point $\om' = \om$ and a large semicircle in the upper half-plane, one obtains the Kramers--Kronig equation
\be F(\om) = \frac{i}{\pi} P \int_{-\infty}^\infty \frac{F(\om')}{\om - \om'} d\om', \label{d31} \ee
where $P$ denotes a principal value \cite{hub89,jac99}. The integral on the right side of Eq. (\ref{d31}), without the factor of $i$, is called a Hilbert transform. By splitting Eq. (\ref{d31}) into real and imaginary parts, one finds that the real and imaginary parts of $F(\om)$ are Hilbert transforms of each other. They contain the same information, so one cannot exist without the other. If $F(t)$ is real, then the real part of $F(\om)$ is an even function of frequency (the cosine transform), whereas the imaginary part is an odd function of frequency (the sine transform). Transforms (\ref{d3}), (\ref{d14}), (\ref{d19}), (\ref{d21}) and (\ref{d23}) all have these properties.

Let $u$ denote the upper limit $\infty$ or $T$. Then it follows from Eqs. (\ref{1.2}) and (\ref{1.3}) that
\ba F(\om) &= &\int_0^u \int_{-\infty}^\infty F(\om') e^{-i\om't} (d\om'/2\pi) e^{i\om t} dt \nonumber \\
&= &\int_{-\infty}^\infty F(\om') \int_0^u e^{i(\om - \om')t} (dt/2\pi) d\om'. \label{d32} \ea
The left and right sides of Eq. (\ref{d32}) are consistent if and only if
\be \int_0^u e^{i(\om - \om')t} dt/2\pi = \de_u(\om - \om'). \label{d33} \ee
The left side of Eq. (\ref{d33}) is the Fourier transform of a step function divided by $2\pi$. Unlike common representations of the $\de$-function, which are real, the natural $\de$-function for one-sided-in-time transforms is complex. It follows from Eqs. (\ref{d14}) and (\ref{d21}) that
\ba \de_\infty(\om) &= &\de(\om)/2 + i/2\pi\om, \label{d34} \\
\de_T(\om) &= &\sin(\om T)/2\pi\om + i[1 - \cos(\om T)]/2\pi\om. \label{d35} \ea
By substituting formula (\ref{d34}) in Eq. (\ref{d32}) and using Eq. (\ref{d31}), one finds that
\ba F(\om) &= &\int_{-\infty}^\infty [\de(\om - \om') + i/\pi(\om - \om')] F(\om') d\om'/2 \nonumber \\
&= &[F(\om) + F(\om)]/2. \ea
Thus, the $\de$-function defined by Eq. (\ref{d34}) sifts any function of frequency that is the infinite-time transform of a causal function. A similar remark applies to the $\de$-function defined by Eq. (\ref{d35}).

\section*{Appendix E: Derivative rule}

One can reconcile the frequency- and time-domain analyses of Secs. 3 -- 5 if one applies the derivative rule carefully. We will prove this statement for an undamped oscillator, because the frequency-domain analysis of this system was the most problematic.

First, recall that the time-domain solution of Eq. (\ref{5.1}) is
\be A(t) = \int_0^t R(s) ds, \label{e1} \ee
from which it follows that
\be \<|A(t)|^2\> = \int_0^t \int_0^t \<R^*(s)R(s')\> ds ds' = \si t. \label{e2} \ee
Second, observe that the Fourier transform of Eq. (\ref{5.1}) is
\be -i\om A(\om) = R(\om) - A(T)e^{i\om T}, \label{e3} \ee
from which it follows that
\be \om^2 \<|A(\om)|^2\> = \<|R(\om)|^2\> -2\re [\<R^*(\om)A(T)e^{i\om T}\>] + \<|A(T)|^2\>. \label{e4} \ee
According to Eqs. (\ref{2.4}) and (\ref{e2}), the first and third terms on the right side of Eq. (\ref{e4}), $\<|R(\om)|^2\>$ and $\<|A(T)|^2\>$, both equal $\si T$. The second term
\ba \<R^*(\om)A(T)e^{i\om T}\> &= &\int_0^T \int_0^T \<R^*(s)R(s')\> e^{i\om(T - s)} ds ds' \nonumber \\
&= &\si (e^{i\om T} - 1)/i\om. \label{e5} \ea
By assembling these results, one finds that
\ba  \om^2 \<|A(\om)|^2\> &= &\si T - 2\si\sin(\om T)/\om + \si T \nonumber \\
&= &2\si T[1 - \sinc(\om T)]. \label{e6} \ea
Equation (\ref{e6}) is equivalent to Eq. (\ref{5.8}). It shows that the terms required to correct Eq. (\ref{5.14}) come from the final value $A(T)$, which is often omitted.

This proof of consistency depends on prior knowledge of the time-domain solution. It is also possible to work entirely in the frequency domain. By setting $\om = 0$ in Eq. (\ref{e3}), one finds that $A(T) = R(0) = \si T$. Hence, one can rewrite Eq. (\ref{e3}) in the alternative form
\be -i\om A(\om) = R(\om) - R(0)e^{i\om T}, \label{e7} \ee
in which the terms on the right side are known.
Equation (\ref{e6}) follows directly from Eq. (\ref{e7}).

\section*{Appendix F: Specific frequency correlations}

Let $R(t)$ be the source function for noise, which has properties (\ref{2.1}) and (\ref{2.2}). For white noise, the two-time correlation
\be \<R^*(t)R(t')\> = \si\de(t - t'). \label{f1} \ee
White noise is $\de$-correlated in time. The associated two-frequency correlation
\ba \<R^*(\om)R(\om')\> &= &\int_0^T \int_0^T \<R^*(t)R(t')\> e^{i(\om't' - \om t)} dt dt' \nonumber \\
&= &\si \int_0^T e^{i(\om' - \om)t} dt \nonumber \\
&= &\si[e^{i(\om' - \om)T} - 1]/i(\om' - \om) \nonumber \\
&= &2\pi\si \de_T(\om' - \om), \label{f2} \ea
where $\de_T$ is the effective $\de$-function that was introduced in App. D. (See Fig. 11.) Thus, white noise is also $\de$-correlated in frequency, as stated after Eq. (\ref{b23}). Notice that interchanging $\om$ and $\om'$ conjugates the correlation. Notice also that $\<R^*(\om)R(\om')\> \rightarrow \si T$ as $\om' \rightarrow \om$. This result is consistent with Eq. (\ref{2.4}).

Now let $A(t)$ be the displacement of a damped oscillator. In Secs. 3 and 4, and App. B, it was shown that $A(\om) \sim G(\om)R(\om)$, where $A(\om)$ and $G(\om)$ are the transformed displacement and Green function, respectively. For such a system, the frequency correlation
\be C(\om,\om') \sim 2\pi\si G^*(\om)G(\om')\de_T(\om' - \om). \label{f3} \ee
Thus, damped oscillators are also (asymptotically) $\de$-correlated in frequency. Notice that $C(\om',\om) = C^*(\om,\om')$.

For colored noise, the temporal correlation
\be \<R^*(t)R(t')\> = (\si\nu/2) e^{-\nu|t - t'|}. \label{f11} \ee
The associated frequency correlation
\be  \<R^*(\om)R(\om')\> = (\si\nu/2) \int_0^T \int_0^T e^{-\nu|t - t'|} e^{i(\om't' - \om t)} dt dt'. \label{f12} \ee
On the right side of Eq. (\ref{f12}), the $t'$-integral is
\ba &&e^{-(\nu + i\om)t} \int_0^t e^{(\nu + i\om')t'} dt' + e^{(\nu - i\om)t}\int_t^T e^{-(\nu - i\om')t'} dt' \nonumber \\
&= &\frac{e^{i(\om' - \om)t} - e^{-(\nu + i\om)t}}{(\nu + i\om')} + \frac{e^{i(\om' - \om)t} - e^{-(\nu - i\om')T} e^{(\nu - i\om)t}}{(\nu - i\om')}. \label{f13} \ea
The $t$-integral of the first term on the right side of Eq. (\ref{f13}) is
\be \frac{e^{i(\om' - \om)T} - 1}{i(\om' - \om)(\nu + i\om')} - \frac{1 - e^{-(\nu + i\om)T}}{(\nu + i\om)(\nu + i\om')} \label{f14} \ee
and the $t$-integral of the second term is
\be \frac{e^{i(\om' - \om)T} - 1}{i(\om' - \om)(\nu - i\om')} - \frac{e^{i(\om' - \om)T} - e^{-(\nu - i\om')T}}{(\nu - i\om)(\nu - i\om')}.\label{f15} \ee
By adding 1 to, and subtracting 1 from, the numerator of the second term in integral (\ref{f15}), and rewriting $i(\om' - \om)$ as $(\nu - i\om) - (\nu - i\om')$, one obtains the contribution
\ba &&\frac{e^{i(\om' - \om)T} - 1}{i(\om' - \om)} \biggl[ \frac{1}{\nu + i\om'} + \frac{1}{\nu - i\om'} - \frac{i(\om' - \om)}{(\nu - i\om)(\nu - i\om')} \biggr] \nonumber \\
&= &\frac{e^{i(\om' - \om)T} - 1}{i(\om' - \om)} \frac{2\nu + i(\om' - \om)}{(\nu + i\om')(\nu - i\om)}. \label{f16} \ea
The remaining contribution is
\be \frac{e^{-(\nu + i\om)T} - 1}{(\nu + i\om)(\nu + i\om')} + \frac{e^{-(\nu - i\om')T} - 1}{(\nu - i\om)(\nu - i\om')}. \label{f17} \ee
Notice that interchanging $\om$ and $\om'$ conjugates contributions (\ref{f16}) and (\ref{f17}). If one neglects the exponentially small terms in the latter contribution, it reduces to
\be -\frac{2(\nu^2 - \om\om')}{(\nu^2 + \om^2)[\nu^2 + (\om')^2]}. \label{f18} \ee
By combining these contributions and multiplying the result by $\si\nu/2$, one obtains the frequency correlation
\be C(\om,\om') = \frac{\si\nu}{2} \Biggl\{ \frac{e^{i(\om' - \om)T} - 1}{i(\om' - \om)} \frac{2\nu + i(\om' - \om)}{(\nu + i\om')(\nu - i\om)} - \frac{2(\nu^2 - \om\om')} {(\nu^2 + \om^2)[\nu^2 + (\om')^2]} \Biggr\}. \label{f19} \ee
Notice that Eq. (\ref{f19}) is consistent with Eq. (\ref{2.10}).

The normalized correlation, which is correlation (\ref{f19}) divided by $\si T$, is illustrated in Fig. 12. Its real part is a symmetric function of $\om$ and $\om'$, whereas its imaginary part is an asymmetric function. The first term on the right side of Eq. (\ref{f19}), which is proportional to the effective $\de$-function, is only significant for a narrow range of frequencies ($|\om' - \om| \sim 1/T$). Although the second term on the right side of Eq. (\ref{f19}) is significant for a wide range of frequencies ($\om$, $\om' \sim \nu$), it is smaller than the first term by a factor of $\nu T$. In this sense, colored noise is $\de$-correlated in frequency.
\begin{figure}[h!]
\vspace{-0.1in}
\centerline{\includegraphics[width=2.8in]{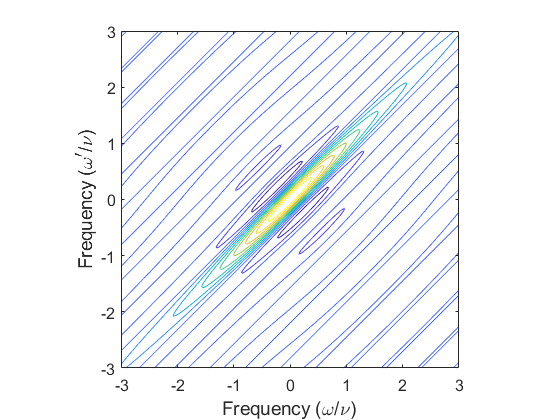} \hspace{-0.4in}
\includegraphics[width=2.8in]{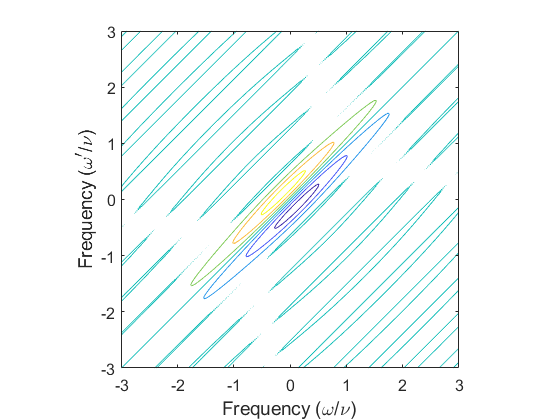}}
\vspace{-0.0in}
\caption{Real part (left) and imaginary part (right) of the frequency correlation of colored noise for $\nu T = 10$. For $|\om' - \om| \gg 1/T$, the correlation is weak. For $\om' = \om$, the correlation is real and equals the spectrum shown in Fig. 1(b).}
\end{figure}

In Sec. 5, it was shown that the temporal correlation of an undamped oscillator is $\si\min(t,t')$. One can calculate the associated frequency correlation by doing the $t$- and $t'$-integrals in Eq. (\ref{b1}) directly. Alternatively, by combining Eqs. (\ref{5.9}) and (\ref{f1}), one finds that 
\ba C(\om,\om') &= &\si \int_0^T \frac{e^{-i\om T} - e^{-i\om s}}{-i\om} \frac{e^{i\om'T} - e^{i\om's}}{i\om'} ds \nonumber \\
&= &(\si/\om\om') \int_0^T \{e^{i(\om' - \om)T} - e^{i\om's - i\om T} \nonumber \\
&&-\ e^{i\om'T - i\om s} + e^{i(\om' - \om)s}\} ds. \label{f21} \ea
The combined integral of the first and fourth terms on the right side of Eq. (\ref{f21}) is
\be Te^{i(\om' - \om)T} + \frac{e^{i(\om' - \om)T} - 1}{i(\om' - \om)}, \label{f22} \ee
whereas the combined integral of the second and third terms is
\be \frac{e^{-i\om T}(1 - e^{i\om'T})}{i\om'} + \frac{e^{i\om'T}(e^{-i\om T} - 1)} {i\om}. \label{f23} \ee
Notice that interchanging $\om$ and $\om'$ conjugates contributions (\ref{f22}) and (\ref{f23}). By combining them and multiplying the result by $\si/\om\om'$, one obtains the frequency correlation
\ba C(\om,\om') &= &\frac{\si}{\om\om'} \Biggl[ Te^{i(\om' - \om)T} + \frac{e^{i(\om' - \om)T} - 1}{i(\om' - \om)} \nonumber \\
&&+\ \frac{e^{-i\om T}(1 - e^{i\om'T})}{i\om'} + \frac{e^{i\om'T}(e^{-i\om T} - 1)} {i\om} \Biggr].  \label{f24} \ea
For very low frequencies ($\om$, $\om' \ll 1/T$), the zeroth- and first-order terms in the square brackets cancel, so the correlation is not singular, and the second-order term is $T^3\om\om'/3$. Hence, $C(\om,\om') \rightarrow \si T^3/3$ as $\om$, $\om' \rightarrow 0$. Notice that Eq. (\ref{f24}) is consistent with Eq. (\ref{5.8}).

The normalized correlation, which is correlation (\ref{f24}) divided by $\si T^3/3$, is illustrated in Fig. 13. Its real part is a symmetric function of $\om$ and $\om'$, whereas its imaginary part is an asymmetric function. If one sections the real and imaginary correlation surfaces along the lines $\om' = -\om$, the curves that result are similar to those in Fig. 11.
Although the heights and depths of the peaks and troughs are independent of $T$, the range of frequencies for which the correlation is significant decreases as $T$ increases. In this sense, the fluctuations of an undamped oscillator are $\de$-correlated in frequency.
\begin{figure}[h!]
\vspace{-0.1in}
\centerline{\includegraphics[width=2.8in]{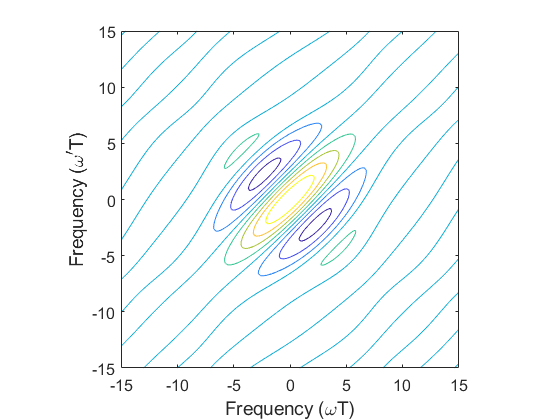} \hspace{-0.4in} \includegraphics[width=2.8in]{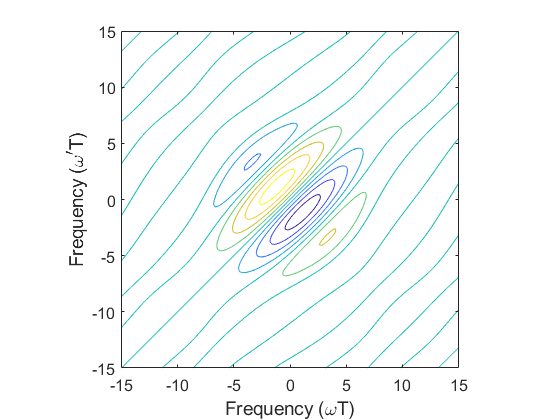}}
\vspace{-0.0in}
\caption{Real part (left) and imaginary part (right) of the frequency correlation of an undamped oscillator. For $|\om' - \om| \gg 1/T$, the correlation is weak. For $\om' = \om$, the correlation is real and equals the spectrum shown in Fig. 8.}
\end{figure}

\end{document}